\documentclass[aps,prb,twocolumn,superscriptaddress,10pt,article,nofootinbib,longbibliography]{revtex4-2}
\usepackage{graphicx,amsfonts,amssymb,amsmath,hyperref,hypcap,enumerate,bm,color}

\newcommand{\ket}[1]{\left|#1\right\rangle}

\newcommand{\beq}{\begin{equation}}
\newcommand{\eneq}{\end{equation}}

\def\qq{\mathbf{q}}
\def\kk{\mathbf{k}}
\def\KK{\mathbf{K}}

\def\KK{\mathbf{K}}
\def\qq{\mathbf{q}}

\begin{document}

\title{Nodal Nematic Superconductivity in Multiple-Flat-Band Land}

\author{Chao-Xing Liu$^*$}
\affiliation{Department of Physics, The Pennsylvania State University, University Park,  Pennsylvania 16802, USA}
\affiliation{Department of Physics, Princeton University, Princeton, NJ 08544} 
\author{B. Andrei Bernevig$^*$} 
\affiliation{Department of Physics, Princeton University, Princeton, NJ 08544} 
\affiliation{Donostia International Physics Center, P. Manuel de Lardizabal 4, 20018 Donostia-San Sebastian, Spain}
\affiliation{IKERBASQUE, Basque Foundation for Science, Bilbao, Spain}

\begin{abstract} 
In this work, we propose a mechanism of inducing stable nodal superconductivity for multiple-flat-band systems. This mechanism is based on the degenerate flat band nature, so that not only intra-eigen-band pairing, but also inter-eigen-band pairing has a significant influence on the superconductivity properties. Based on the Bogoliubov de Gennes formalism of the heavy fermion model for the twisted bilayer graphene as an example, we show that although the nodal nematic d-wave pairing has higher energy around the nodal points compared to the chiral d-wave pairing in the momentum space, the inter-eigen-band pairing can lower the energy in other momenta away from the nodes, so that the nematic d-wave pairing is energetically favored for its overall condensation energy. This type of mechanism is particular to multiple-flat-band systems with no Fermi surfaces. 
\end{abstract}

\date{\today}


\maketitle

{\it Introduction - } 
In the Bardeen-Cooper-Schrieffer (BCS) theory, when electrons experience attractive interactions mediated by phonons, a superconductor (SC) pairing instability can arise, leading to a gap opening at the electron Fermi surface \cite{tinkham2004introduction}. Typically, the Fermi surface becomes fully gapped to lower the SC ground state energy, resulting in nodeless s-wave superconductivity. In contrast, nodal superconductivity can emerge when the SC pairing is of an "unconventional" nature \cite{sigrist1991phenomenological}. The most notable example is the high $T_c$ superconductivity in cuprates \cite{tsuei2000pairing}, where the d-wave nature of SC pairing leads to nodes at the Fermi surface. The occurrence of nodal superconductivity usually requires unconventional SC pairing mechanism or wavefunction "topology", beyond the phonon mediated superconductivity in normal metals in the BCS theory, as the nodes at the Fermi surface are not energetically favored. However, this scenario might be changed for superconductivity in multiple-flat-bands systems when the band width and energy separation between different flat bands are smaller than the attractive interactions between electrons (strong coupling limit), even for the phonon-mediated SC systems, which is the main topic of this work. This topic could also be relevant for the Fermi surfaces with topological obstructions in the presence of electron-phonon interaction. 

Flat-band superconductivity has attracted great research interests due to the discovery of superconductivity in Moir\'e SC materials, e.g twisted bilayer/trilayer graphene \cite{cao2018unconventional,cao2018correlated}. 
While the microscopic origin of superconductivity in these Moir\'e SC materials still remains elusive \cite{khalaf2021charged,khalaf2022symmetry,po2018origin,you2019superconductivity,lian2019twisted,wu2018theory,wu2019topological,wu2019phonon,wu2019identification,chichinadze2020nematic,fernandes2021charge,wang2021topological,yu2022euler2,sharma2020superconductivity,isobe2018unconventional,roy2019unconventional,peltonen2018mean,chichinadze2020nematic,kennes2018strong,gonzalez2019kohn,cea2021coulomb,shavit2021theory,liu2018chiral,fidrysiak2018unconventional,sharma2020superconductivity,chou2019superconductor,kozii2019nematic,you2019superconductivity,wang2021topological,hu2019geometric,wagner2024coulomb,wang2024molecular,wang2024electron,shi2024moir,christos2023nodal}, the experimental studies of the SC gap in twisted bilayer graphene (TBG) suggest an unconventional nature of SC pairing, beyond the conventional s-wave spin-singlet pairing in the BCS theory. 
The scanning tunneling microscope (STM) \cite{oh2021evidence} and electronic thermal conductance measurements \cite{di2022revealing} on TBG suggested a nodal SC pairing gap; evidence of nematicity has also been found in the SC state via magneto-transport measurements \cite{cao2021nematicity}. On the other hand, recent ARPES experiments with high spatial resolutions found the replicas of flat bands that are expected to result from the strong electron-$K$-phonon interaction \cite{chen2023strong}. This raises the academic questions if phonon-mediated superconductivity can reveal the unconventional features such as nodes and nematicity in the SC state, which can be more stable than the fully gaped SC state, in flat-bands land. Note that we are not claiming this as the TBG superconductivity mechanism due to the strong Coulomb interaction, which needs to be screened for SC to appear. 

In Ref.\cite{liu2024electron}, we have shown that the electron-phonon interaction between the bare bands of TBG and the K-phonons \cite{kwan2023electron,shi2024moir} can induce two types of pairing channels, the intra-Chern-band and inter-Chern-band pairings, based on the heavy fermion model \cite{song2022magic,shi2022heavy,herzog2024topological}. The intra-Chern-band pairing is d-wave and spin singlet, and belongs to the 2D $E_2$ irreducible representation (irrep) of the $C_{6v}$ group. Strikingly, an Euler obstructed pairing that belongs to the intra-Chern-band pairing has recently been proposed to understand the nematicity and nodal superconductivity in TBG \cite{yu2022euler2,wang2024molecular}. Even though the SC nature in TBG remains elusive,  
the academic question of the nematic and nodal SC is relevant as the evidence of these features have been found in TBG experiments \cite{oh2021evidence,di2022revealing,cao2021nematicity}. Due to the 2D irrep nature, other intra-Chern-band pairing channels, e.g. chiral d-wave pairing, can also exist and share the same $T_c$ as the Euler obstructed pairing. Naively, one may anticipate chiral d-wave pairing is more stable at zero temperature as it is fully gapped while the nematic Euler pairing has nodes. Such situation occurs in the doped graphene model \cite{nandkishore2012chiral,black2014chiral}. Thus, it is interesting to theoretically understand the stability of nematic Euler pairing with nodes.

In this work, we study the low temperature behavior of the 2D $E_2$-irrep intra-Chern-band pairing induced by the electron-$K$-phonon interaction in the TBG heavy fermion model\cite{song2022magic,shi2022heavy}. Our numerical results of the condensation energy suggest the nematic Euler pairing has a lower energy than the chiral d-wave pairing and has nodes for a wide range of parameters in the chiral limit,
although this limit is away from experimental situations. A detailed theoretical analysis shows the nematic Euler pairing is at least locally stable. The underlying mechanism is that although gapping out the nodes in Euler pairing can lower energy around the nodal regions in the momentum space, strong inter-eigen-band pairing potential can lead to an energy saving for other momenta away from the nodal regions, which is the dominant contribution. This mechanism is proposed to represent a counter example to the known fact that gapping gapless points leads to lower energy. 

{\it Bogoliubov-de gennes Hamiltonian for the flat bands of TBG -}
Our previous work \cite{liu2024electron} has shown that in absence of Coulomb interaction, the coupling between the flat bands and the $K$-phonons in TBG model can potentially favor two pairing channels, the s-wave spin-singlet inter-Chern-band pairings with 1D $A_1$ irreducible representation (irrep) of $C_{6v}$ group and the d-wave spin-singlet intra-Chern-band pairings with the 2D $E_2$ irrep. These two channels are decoupled in the chiral limit and their $T_c$'s are close (for the bare flat bands). 
We focus on the d-wave intra-Chern-band channel below. Due to the 2D $E_2$ irrep nature, all different forms of the d-wave intra-Chern-band pairings share the same $T_c$. The degeneracy of different 2D $E_2$ irrep pairing channels will be broken down below $T_c$, which can be extracted from the full gap equation. To study that, we consider the Bogoliubov-de Gennes (BdG) Hamiltonian of TBG with spin singlet pairing, 
\begin{eqnarray}
&\mathcal{H}^{\eta}_{BdG} =\frac{1}{2} \sum_{\kk\in MBZ} \psi_{\kk, \eta}^\dagger H_{BdG}^\eta(\kk) \psi_{\kk, \eta},\;\;\;\nonumber \\
& H_{BdG}^{\eta=+}(\kk) =\begin{pmatrix}
h_+(\kk)\otimes s_0 & 2\Delta_{\kk} \otimes (i s_y) \\
2\Delta_{\kk}^\dagger\otimes  (-i s_y) & - h_{-}^\star(-\kk) \otimes s_0
\end{pmatrix}\label{eq:HBdGp} \\ &  H_{BDG}^{\eta=-}(\kk) =\begin{pmatrix}
h_-(\kk)\otimes s_0 & 2\Delta_{-\kk}^T \otimes (i s_y) \\
-2\Delta_{-\kk}^\star \otimes  (i s_y) & - h_{+}^\star(-\kk) \otimes s_0
\end{pmatrix},\nonumber 
\end{eqnarray} 
where $\eta=\pm$ is the valley index and $s_0$ and ${\bf s}$ label the identity and Pauli matrices for spin. 
The basis functions are $\psi_{\kk, \eta}^\dagger = (\gamma^\dagger_{\kk, \eta, e_Y = \pm ,s =\uparrow \downarrow},\gamma_{-\kk, -\eta , e_Y=\pm,s=\uparrow \downarrow})$, where $\gamma^\dagger_{\kk, e_Y ,\eta,s}$ is the electron creation operator for the flat bands in TBG with the Chern-band index $e_Y=\pm$ and spin $s$. Here the Chern-band basis is related to the eigen-state basis by
\begin{equation}\label{eq:Chern-band-basis}
\gamma^{\dag}_{\mathbf{k}, \eta, e_Y, s}=\frac{1}{\sqrt{2}}(\gamma^\dag_{\mathbf{k},\eta, n=+, s} + i e_Y \gamma^\dag_{\mathbf{k}, \eta, n=- , s})
\end{equation}
for the projected single-particle Hamiltonian, where $\gamma^\dag_{\mathbf{k},\eta, n, s}$ is the electron creation operator for the eigen-state $n=\pm$ of two flat bands per valley per spin. We use the Chern-band basis to write down the BdG Hamiltonian because the inter-Chern-band and intra-Chern-band pairings are decoupled in the chiral limit \cite{ledwith2020fractional,tarnopolsky2019origin,liu2024electron}. We note that the realistic TBG is away from the chiral limit and thus our discussion is of academic interest. On the Chern-band basis, the projected single-particle Hamiltonian on the flat bands at the valley $\eta$ reads
\beq h_\eta(\kk)= (d_{0,\eta}(\kk)-\mu)\zeta^0+d_{x,\eta}(\kk)\zeta^x, \eneq
where $\mu$ is the chemical potential, $\zeta^0$ and $\zeta^{x,y,z}$ represent the identity matrix and Pauli matrices for the Chern-band basis, $d_{0,\eta}(\kk)=(\epsilon_{+,\eta}(\kk)+\epsilon_{-,\eta}(\kk))/2; d_{x,\eta}(\kk)=(\epsilon_{+,\eta}(\kk)-\epsilon_{-,\eta}(\kk))/2, $
and $\epsilon_{n, \eta}(\kk)$ are the single-particle eigen-energies that can be extracted from the Bistritzer-MacDonald (BM) model \cite{bistritzer2011moire,liu2024electron}. 
Time reversal (TR) symmetry requires $h_{-\eta}(-\kk)=h_\eta(\kk)$. We always consider the chiral limit with $d_{0,\eta}(\kk)=0$ below. The intra-Chern-band pairing has the form
\beq \Delta_{\kk} = \left(\begin{array}{cc} \Delta_{\kk,+}&0\\
0 & \Delta_{\kk,-}\end{array}\right), \eneq
where $\Delta_{\kk,e_Y}$ labels the gap function on the Chern-band channel $e_Y=\pm$. The intra-Chern-band pairing gap function $(\Delta_{\kk,+},\Delta_{\kk,-})$ belong to the 2D $E_2$ irrep and has a d-wave nature, which can be represented by \cite{liu2024electron}
\beq \label{eq:dwave_ansatz} \Delta_{\kk,e_Y} = \Delta_{e_Y} \frac{k^2_{-e_Y}}{k^2+b^2} \eneq
with $k_{-e_Y}=k_x-i e_Y k_y$. In particular, $(\Delta_{+},\Delta_{-})= (\Delta_0,0)$ or $ (0,\Delta_0)$ gives the TR-breaking chiral d-wave while $(\Delta_{+},\Delta_{-})=(\Delta_0 , \Delta_0^* )$ represents the TR-preserving nematic Euler pairing. $H_{BdG}^-(\kk)$ can be related to $H_{BdG}^+(\kk)$ by SC particle-hole symmetry (See Appendix Sec. A), so we focus on $H_{BdG}^+(\kk)$, which has a block diagonal form in the spin space (see Eq. \ref{eq:HBdGp}) and we can further decompose it into $H_{BdG}^+(\kk)=H_{BdG}^{+,+}(\kk)+H_{BdG}^{+,-}(\kk)$ with $H_{BdG}^{+,\lambda}$ a 4-by-4 matrix Hamiltonian (See Eq. A10 in Appendix Sec. A for $H_{BdG}^{+,\lambda}$). The eigen-energy of $H_{BdG}^{+,\lambda=+}$ can be solved analytically as 
\begin{eqnarray} \label{eq:BdGEnergy1}
& E^{\pm}_{\kk,n} = \pm \frac{1}{2}\sqrt{2A+\mu^2+d_{x,\kk}^2 + 2 n \sqrt{L_\kk }}
\end{eqnarray}
where $d_{x,\kk}=d_{x,\eta=+}(\kk)$, $n=\pm$
and
\beq L_\kk = A^2-4B^2+\mu^2d^2_{x,\kk}+d_{x,\kk}^2(A-2B \cos(\Phi_{\kk})) \nonumber
\eneq
with $A=|\Delta_{\kk,+}|^2+|\Delta_{\kk,-}|^2$, $B=|\Delta_{\kk,+}\Delta_{\kk,-}|$ and $\Phi_\kk=\phi_{\kk,-}-\phi_{\kk,+}$ with $\phi_{\kk,e_Y}=\Delta_{\kk,e_Y}/|\Delta_{\kk,e_Y}|$. As $h_{\eta}(\kk)$ is not diagonal in the Chern-band basis, the nodes of the BdG Hamiltonian do not coincide with $\Delta_{\kk,e_Y}=0$ that would only occur at $\kk=0$ for non-zero $\Delta_{e_Y}$. The Euler pairing is defined by $|\Delta_{+}|=|\Delta_{-}|$\cite{yu2022euler1,yu2022euler2}, which possesses the nodes at $\kk$ satisfying $4 |\Delta_{\kk,+}\Delta_{\kk,-}| + \mu^2=d_{x,\kk}^2$ and $\cos\left(\frac{\Phi_{\kk}}{2}\right)=0$. The general conditions for the nodes are discussed in Ref. \cite{liu2024electron} and Appendix Sec. A. 
The eigen-states of $H_{BdG}^{+,+}$ can be encoded in the unitary matrix $U_\kk$ defined by
\beq \label{eq:UMatrix_1}
U_\kk =\begin{pmatrix}
u_\kk & v_\kk \\
w_{\kk} & r_{\kk} 
\end{pmatrix}, \;\; H_{BdG}^{+,+}(\kk)= U_\kk D_\kk U^\dagger_\kk
\eneq
where $u_\kk, v_\kk, w_{\kk}, r_{\kk}$ are two-by-two matrices and $D_\kk$ is diagonal, $D_\kk=\text{Diag}(E^+_+,E^+_-,E^-_+,E^-_-)$, with $E^+_{n}>0$ and $E^-_{n}<0$ ($n=\pm$). 


\begin{figure*}[hbt!]
   \centering
    \includegraphics[width=7in]{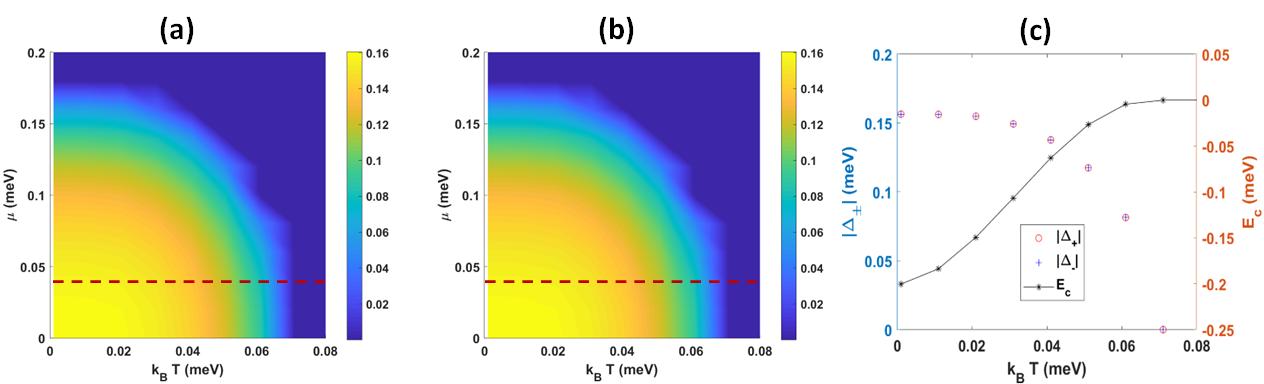} 
    \caption{ The self-consistent gap functions (a) $|\Delta_+|$ and (b) $|\Delta_-|$ as a function of $k_BT$ and $\mu$ in the chiral limit. (c) For fixed chemical potential $\mu=0.04meV$ (along the red dashed line in (a) and (b)), the gap function $|\Delta_{\pm}|$ and the condensation energy $E_c$ are plotted as a function of temperature $k_B T$. 
    }
    \label{fig1:PhaseDiagram}
\end{figure*}

{\it Gap equation and Superconductor Phase Diagram - }
To derive the self-consistent gap equation below $T_c$, we consider the attractive electron-electron interaction mediated by K-phonons \cite{liu2024electron}, 
\begin{eqnarray}\label{eq:Interactionform_intra_Chern1}
V^{\eta,e_Y,e_Y}_{\kk,\kk'}=U^*_{1,e_Y,\kk} U_{1,e_Y,\kk'}; \; U_{1,e_Y,\kk}=\frac{\sqrt{V_0}}{k^2+b^2} k_{e_Y}^2;  
\end{eqnarray}
where $k_{e_Y}=k_x+i e_Y k_y=k e^{i e_Y \theta_\kk}$, $e_Y=\pm$, and $V_0=0.4 meV$ and $b=0.185 k_\theta$ are material dependent parameters given in Ref.\cite{liu2024electron} (Also see Appendix Sec. A).
With the above form of $V$-interaction and gap function ansatz in Eq. (\ref{eq:dwave_ansatz}), the self-consistent gap equation is 
\begin{eqnarray} \label{eq:GapEquation_intra_dwave1} 
&& \Delta_{e_Y}= \frac{V_0 }{N_M} \sum_{\kk', n } \frac{{k'}^2_{e_Y}}{k'^2+b^2} \left( u_{-\kk',e_Y n} w^*_{-\kk',e_Y n} \right. \nonumber \\
&& \left. (1- f(E^+_{-\kk',n}))  + v_{-\kk',e_Y n} r^*_{-\kk', e_Y n} f(-E^-_{-\kk',n}) \right), 
\end{eqnarray}
with $N_M$ the number of moir\'e unit cells. Using the gap equation (\ref{eq:GapEquation_intra_dwave1}) at zero temperature, the condensation energy is given by
\beq \label{eq:condensation_energy_2} 
E_c=4 \left( \sum_{\kk, n} (E^-_{\kk,n}-\xi^-_{\kk,n})  + \frac{ N_M}{V_0} \sum_{e_Y } |\Delta_{e_Y}|^2 \right),  \eneq
where $\xi^-_{\kk,n}=-\frac{1}{2}\left|d_{0,\kk}-\mu+n d_{x,\kk}\right|$ is the eigen-energy of the single-particle Hamiltonian.

The SC phase diagram can be extracted by solving the self-consistent gap equation (\ref{eq:GapEquation_intra_dwave1}). 
Fig. \ref{fig1:PhaseDiagram} (a) and (b) show the amplitude of the gap functions $\Delta_+$ and $\Delta_-$ as a function of the chemical potential $\mu$ and the temperature $k_B T$. These two phase diagrams are almost identical, suggesting that the nematic Euler pairing with $|\Delta_+|=|\Delta_-|$ is the self-consistent solution of the gap equation (\ref{eq:GapEquation_intra_dwave1}). Fig. \ref{fig1:PhaseDiagram} (c) shows more details of the gap functions and the corresponding condensation energy as a function of $k_B T$ for $\mu=0.04 meV$ (along the red dashed lines in Fig. \ref{fig1:PhaseDiagram} (a) and (b)). 

\begin{figure}[hbt!]
  \centering
    \includegraphics[width=3.5in]{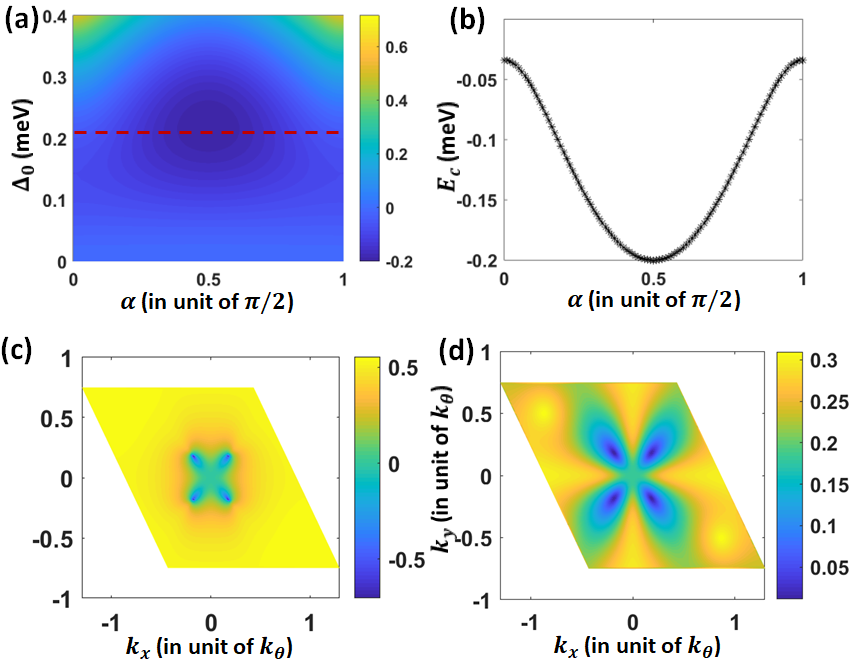} 
    \caption{ (a) The condensation energy $E_c$ is plotted as a function of $\alpha$ and $\Delta_0$ for $\phi=0$. (b) $E_c$ is plotted as a function of $\alpha$ for $\Delta_0\approx 0.22$ meV (the red dashed line in (a)). The minimum of $E_c$ appears at $\alpha=\pi/4$ (Euler pairing). (c) $\beta_\kk$ is plotted as a function of $\kk$ in MBZ. $\beta_\kk$ is defined in Eqs. (\ref{eq:Ec_expansion}) and (\ref{eq:betakk_expansion}). (d) The energy gap of the BdG spectrum is plotted as a function of $\kk$ for $\alpha=\pi/4$. The energy unit here is meV.     }
    \label{fig2:EnergyDiff}
\end{figure}

To investigate the stability of the nematic Euler pairing, we consider a general form of the gap function,
\beq \label{eq:gapfunction_ansatz} (\Delta_+,\Delta_-)=\Delta_0 (\cos\alpha e^{-i\phi/2},\sin\alpha e^{i\phi/2}), \eneq
where $\Delta_0$ is real, the parameter $\alpha$ tuning the gap function form in 2D $E_2$ irrep and $\phi$ is the relative phase. The nematic Euler pairing corresponds to $\alpha=\frac{\pi}{4}$ while the chiral d-wave pairing corresponds to $\alpha=0$ or $\frac{\pi}{2}$. The BdG spectrum is generally gaped unless $\alpha=\pi/4$. $E_c$ has a very weak dependence on $\phi$ (See Fig. S3(B) and the corresponding discussion in Appendix Sec. B), and thus we choose $\phi=0$ as an example to show the $\Delta_0$ and $\alpha$ dependence of $E_c$ in Fig. \ref{fig2:EnergyDiff} (a), in which the red dashed line shows the minimal value of $E_c$ located at $\Delta_0\approx 0.22$ meV. This value of $\Delta_0$ is consistent with that obtained from the self-consistent gap equation (\ref{eq:GapEquation_intra_dwave1}) ($\Delta_0^2=|\Delta_+|^2+|\Delta_-|^2$). 
Fig. \ref{fig2:EnergyDiff} (b) shows $E_c$ as a function of $\alpha$ at $\Delta_0\approx 0.22$ meV, in which the minima of $E_c$ appears at $\alpha=\pi/4$. These numerical results suggest the Euler pairing has a lower energy than the chiral d-wave pairing with $\alpha=0$ and is at least locally stable. We will next provide a theoretical understanding of this local stability.

{\it Locally Stable Nematic Euler Pairing - }
To understand this local stability, we consider the gap function slightly away from the Euler pairing and study the variation of the condensation energy at zero temperature. Particularly, we choose $\alpha=\frac{\pi}{4}+\delta_0$ with $\delta_0\ll 1$ in the gap function ansatz (\ref{eq:gapfunction_ansatz}) and expand the condensation energy $E_c$ in Eq. (\ref{eq:condensation_energy_2}) as 
\beq \label{eq:Ec_expansion} 
E_c=E_{c0}+\gamma \delta_0 +\beta \delta^2_0  \eneq
up to the $\delta_0^2$ order. Our main result below is to demonstrate $\gamma=0$ and $\beta>0$ so that Euler pairing becomes locally stable. Furthermore, we define 
\beq \label{eq:betakk_expansion} \beta=\sum_\kk \beta_\kk, \eneq
and the momentum distribution of $\beta_\kk$ is shown in Fig. \ref{fig2:EnergyDiff}(c), from which one can see that the negative value of $\beta_\kk$ around the nodal regions is compensated by the positive $\beta_\kk$ away from the nodal regions. 

Now let us discuss different terms of $E_c$ in Eq. (\ref{eq:condensation_energy_2}). The term $\frac{ N_M}{V_0} \sum_{e_Y }|\Delta_{e_Y}|^2$ in $E_c$ is independent of $\delta_0$, and thus only contributes to $E_{c0}$. The expansion of the $\sum_{\kk, n} (E^-_{\kk,n}-\xi^-_{\kk,n}) $ term in $E_c$ is subtle due to the existence of the nodes in the BdG spectrum. The location of the nodes in the momentum space is denoted as $\kk_g$, whose values are determined by the conditions 
\beq \label{eq:gapless_condition} \cos(\Phi_{\kk_g}/2)=0; d_{x,\kk_g}^2=\mu^2+2\Delta_{0}^2\frac{k_g^4}{(k_g^2+b^2)^2}. \eneq
We divide the Moir\'e Brillouin zone (MBZ) into two regions: one region is around the point node $\kk_g$, denoted as $\Omega_{\kk_g}$, namely $\kk=\kk_g+\qq$ with $|\qq|<\varepsilon$, where $\varepsilon$ serves as a momentum cut-off, and the other is the momentum region away from the point nodes, namely $\kk \not\in \Omega_{\kk_g}$. 

In the region $\kk\in \Omega_{\kk_g}$, we can expand the BdG eigen-energy for the $n=-$ branch around the nodal points and find
\begin{eqnarray} \label{eq:Emkkm_1}
&& E^-_{\kk,-} \approx -\sqrt{ v_1^2q_1^2+v_2^2q_2^2+ B_3\delta_{0}^2 },  
\end{eqnarray}
for small $\qq$ and $\delta_0$, where $q_1=\frac{1}{\sqrt{2}}(q_x+q_y)$, $q_2=\frac{1}{\sqrt{2}}(q_x-q_y)$, and $v_{1,2}$ and $B_3$ are the parameters extracted from the perturbation expansion (See Appendix Sec. C). In Eq. (\ref{eq:Emkkm_1}), $E^-_{\kk,-}$ is gapless at $\qq=0, \delta_0=0$. Although a small but finite $\delta_0$ will give a gap that is linearly proportional to $|\delta_0|$, we find its contribution to the condensation energy $E_c$ requires the momentum $\qq$ integral of Eq. (\ref{eq:Emkkm_1} in the region $\Omega_{\kk_g}$, and this integral gives the term of order $\delta^2_0$ (see Appendix Sec.C) under the condition
\beq \label{eq:q-cut-off} |B_3 \delta_0|\ll |\bar{v} \varepsilon|, \eneq
where $\bar{v}=\frac{v_1+v_2}{2}$ is the average velocity around the point nodes. In our numerical calculations, we first choose the cut-off $\varepsilon$ so that the energy dispersion (\ref{eq:Emkkm_1}) is valid in $\Omega_{\kk_g}$ and then choose $\delta_0$ to satisfy Eq. (\ref{eq:q-cut-off}). 
The correction to $\beta$ in the region $\Omega_{\kk_g}$ is approximately given by (See Appendix Sec. C)
\begin{eqnarray} \label{eq:beta_sumk_inOmega}
\sum_{\kk \in \Omega_{\kk_g}}\beta_\kk \approx - \frac{\pi \varepsilon^2  S_M}{(2\pi)^2} \frac{4 B^2_3 }{ |\bar{v}| \varepsilon } , 
\end{eqnarray}
where $S_M$ is the area of the Moir\'e unit cell. The negative sign here suggests that this contribution will lower $E_c$ for a finite $\delta_0$. 
In numerical calculations, we choose $\varepsilon \approx 0.01 k_\theta$, so that the area ratio between $\Omega_{\kk_g}$ and MBZ, represented by the factor $\frac{\pi \varepsilon^2  S_M}{(2\pi)^2}$ in Eq. (\ref{eq:beta_sumk_inOmega}), is only around $10^{-4}$, and the energy dispersion within $\Omega_{\kk_g}$ is well described by Eq.(\ref{eq:Emkkm_1}). With this value of $\varepsilon$, we then set $\delta_0=0.01\times \frac{\pi}{2}\approx 0.0157$ to satisfy Eq. (\ref{eq:q-cut-off}) as $\varepsilon \approx 0.01 k_\theta \gg \frac{|B_3\delta_0|}{|\bar{v}|} \approx 0.0014 k_\theta$ with $\bar{v} \approx  0.4 meV/k_\theta$ and $B_3 \approx 0.036 meV$. The contribution to $\beta$ from the $\Omega_{\kk_g}$ region is estimated as $\sum_{\kk \in \Omega_{\kk_g}}\beta_\kk  \approx - 1.6\times 10^{-4}$ meV. 


For the $n=+$ branch in the whole MBZ or the $n=-$ branch in the momentum region $\kk \not\in \Omega_{\kk_g}$, the BdG spectrum has a gap, so we can perform the perturbation expansion for $E^-_{\kk,n}$ for a small $\delta_0$, 
\begin{eqnarray}\label{eq:Eng_expansion}
E^-_{\kk,n} \approx -f_{0\kk,n}-\frac{f_{2\kk,n}}{2f_{0\kk,n}}\delta_{0}^2 
\end{eqnarray}
up to the order of $\delta_0^2$, where 
\begin{eqnarray} \label{eq:f0_eigenenergy}
&& f_{0\kk,n}=\frac{1}{2} \left[\left(|d_{x,\kk}|+n\sqrt{\mu^2+2\Delta_{0\kk}^2\sin^2(\Phi_\kk/2)}\right)^2 \right. \nonumber \\
&& \left. + 2\Delta_{0\kk}^2\cos^2(\Phi_\kk/2) \right]^{1/2},
\end{eqnarray}
and
\beq f_{2\kk,n}= \frac{n}{2} \frac{\Delta_{0\kk}^2(2\Delta_{0\kk}^2+d_{x,\kk}^2\cos(\Phi_\kk))}{|d_{x,\kk}|\sqrt{\mu^2+2\Delta_{0\kk}^2\sin^2(\Phi_\kk/2)}} \eneq
for $n=\pm$. From the above derivation, one can see that the lowest order contribution to $E_c$ is of order $\delta_0^2$, no matter inside or outside $\Omega_{\kk_g}$, so $\gamma=0$ in Eq. (\ref{eq:Ec_expansion}). The full form of $\beta$ in Eq. (\ref{eq:Ec_expansion}) is given by
\begin{eqnarray}\label{eq:beta_full}
\beta =   - \frac{\pi \varepsilon^2  S_M}{(2\pi)^2} \frac{4 B^2_3 }{ \bar{v} \varepsilon } - \sum_{n,\kk\not\in \Omega_{\kk_g}} \frac{2f_{2\kk,n}}{f_{0\kk,n}}, 
\end{eqnarray}
in which the first and second terms are the contributions inside and outside $\Omega_{\kk_g}$, respectively. We numerically find $\beta_{\kk \not\in \Omega_{\kk_g}}$ of order $0.1$ meV, several order larger than the contribution inside $\Omega_{\kk_g}$ ($\sim 10^{-4}$ meV). Thus, it is safe to neglect the first term in $\beta$. We numerically evaluate $\beta_\kk$ from the second term in Eq. (\ref{eq:beta_full}), which is shown in Fig. \ref{fig2:EnergyDiff}(c). By comparing to the BdG energy gap shown in \ref{fig2:EnergyDiff}(d), we find a negative $\beta_\kk$ around the point nodes, while $\beta_\kk$ for a large momentum is always positive. All these features of $\beta_\kk$ can be understood from Eq. (\ref{eq:beta_full}) (See Appendix Sec. C). After summing over the momenta for $\beta_\kk$, our numerical results show that the positive $\beta_\kk$ for large momenta is dominant and $\beta\approx 0.44 meV > 0$. This concludes the Euler pairing for $\alpha=\pi/4$ is locally stable. 


The locally stable nematic Euler pairing in our model is in sharp contrast to the case of doped graphene model\cite{nandkishore2012chiral,black2014chiral}, in which the chiral d-wave pairing with a full gap is found to be more stable. Unlike the dispersive bands in doped graphene, both the bandwidth and the energy splitting between different bands are small for two flat bands (per valley per spin) in the TBG model. Thus, the pairing potential not only exists within one eigen band (intra-eigen-band pairing), but also can strongly mix different eigen-bands (inter-eigen-band pairing). This inter-eigen-band mixing due to the pairing term is much smaller than the energy splitting between two bands in doped graphene. 
To see the influence of the inter-eigen-band pairing, we manually turn off the inter-eigen-band pairing term and perform a similar analysis of $E_c$ expanded around $\alpha=\pi/4 + \delta_0$ ($\delta_0\ll 1$) in Appendix Sec.D. Our numerical calculation gives a negative $\beta$ in the $\delta_0^2$ term in Eq. (\ref{eq:Ec_expansion}), which implies that the Euler pairing becomes unstable. 
Fig. S6 in Appendix Sec.D shows the minima of $E_c$ in this case is deviated from the Euler pairing with $\alpha=\pi/4$. 


{\it Conclusion - }
This work suggests that the gapless nematic Euler pairing can be locally stable in the nearly degenerate flat band systems as the inter-eigen-band pairing potential, which is comparable to the bandwidth of each band and band gap between two flat bands, can induce a strong inter-band mixing. Although our discussion focused on the specific model of TBG, the inter-band-pairing mechanism of nodal superconductivity may also exist in other multiple-flat-bands SC systems.

{\it Acknowledgement} --
We would like to acknowledge Y. Chen and A. Yazdani for the helpful discussion. BAB was primarily supported by DOE Grant No. DESC0016239, and also acknowledged the sabbatical support provided by the Simons Investigator grant (No. 404513) and the European Research Council (ERC) under the European Union's Horizon 2020 research and innovation programme (grant agreement no. 101020833). CXL acknowledges the support from the NSF-MERSEC (Grant No. MERSEC DMR 2011750) and the support in part by grant NSF PHY-1748958 to the Kavli Institute for Theoretical Physics (KITP).

\bibliographystyle{apsrev4-2}
\bibliography{ref_epTBG}

\appendix 
\newpage

\renewcommand{\thefigure}{S\arabic{figure}}

\setcounter{figure}{0}

\onecolumngrid


\section{Model Hamiltonian for stable gapless nematic superconductivity}
Some of derivations in this section has been presented in Ref. \cite{liu2024electron}, but in order to keep the current work self-contained, we will still present a systematic derivation of the possible superconducting (SC) channels of TBG induced by $K$-phonons in this section. 

There are two pairing channels that are potentially favored by $K$-phonons, the d-wave spin-singlet intra-Chern-band pairings with the 2D $E_2$ irreducible representation (irrep) and the s-wave spin-singlet inter-Chern-band pairings with 1D $A_1$ irrep. The two channels are decoupled in the chiral limit\cite{ledwith2020fractional,tarnopolsky2019origin,liu2024electron}. Here we are interested in the intra-Chern-band channel. Due to the 2D $E_2$ irrep nature, all different forms of the d-wave intra-Chern-band pairings will share the same $T_c$, and thus it is natural to ask what is the exact form of intra-Chern-band pairing for the temperature below $T_c$, particularly at zero temperature. To study that, we consider the interacting Hamiltonian
\beq \mathcal{H}_{tot}= \mathcal{H}_0 + \mathcal{H}_{el-el} \eneq
where 
\beq \mathcal{H}_0= \sum_{\kk e_1 e_1 s \eta} \gamma^\dagger_{\kk,e_1, \eta, s} h_{\eta, e_1 e_2}(\kk)\gamma_{\kk, e_2, \eta, s} \eneq
and 
\beq \label{eq:Elel_interaction_ChernBasis}
\mathcal{H}_{el-el}=-\frac{1}{N_M} \sum_{\kk,\kk',s_1, s_2, e_Y, e_Y'} V^{\eta, e_Y, e_Y'}_{\kk, \kk'} 
\gamma_{\kk,e_Y,\eta, s_1}^\dagger \gamma_{-\kk, e_Y', - \eta, s_2}^\dagger \gamma_{-\kk', e_Y', \eta, s_2} \gamma_{\kk', e_Y,-\eta, s_1}. 
\eneq
$\mathcal{H}_0$ is the non-superconducting effective Hamiltonian of the flat bands on the Chern-band basis and the form of $h_{\eta=\pm}(\kk)$ can be extracted from the Bistritzer-MacDonald (BM) model \cite{bistritzer2011moire} with twist angle $\theta=1.06^\circ$ and is written as
\beq h_\eta(\kk)= (d_{0,\eta}(\kk)-\mu)\zeta^0+d_{x,\eta}(\kk)\zeta^x, \eneq
where $\mu$ is the chemical potential, $\eta$ is the valley index, $\zeta^0$ and $\zeta^{x,y,z}$ represent the identity matrix and Pauli matrices for the Chern-band basis, $d_{0,\eta}(\kk)=(\epsilon_{+,\eta}(\kk)+\epsilon_{-,\eta}(\kk))/2; d_{x,\eta}(\kk)=(\epsilon_{+,\eta}(\kk)-\epsilon_{-,\eta}(\kk))/2, $
and $\epsilon_{n, \eta}(\kk)$ ($n=\pm$) are the single-particle eigen-energies. $d_{0,\eta}(\kk)$ is generally non-zero in the BM model \cite{bistritzer2011moire}, while the chiral limit \cite{ledwith2020fractional,tarnopolsky2019origin} will require $d_{0,\eta}(\kk)=0$. 
Time reversal symmetry requires $h_{-\eta}(-\kk)=h_\eta(\kk)$. 

$\mathcal{H}_{el-el}$ describes the attractive electron-electron interaction mediated by the K-phonons with the interaction function $V^{\eta, e_Y, e_Y'}_{\kk, \kk'}$ written on the Chern-band basis $e_Y$ and $e'_Y$. Under the mean-field decomposition with the gap function defined by 
\beq \label{eq:Decomposition_GapFunctionDef}
 \Delta^{\eta}_{\kk; e_{1} s_1, e_2 s_2}= - \frac{1}{N_M}  \sum_{\kk'} V_{\kk\kk'}^{\eta e_1 e_2} \langle \gamma_{-\kk'  e_2 \eta s_2} \gamma_{\kk' e_1 -\eta s_1}\rangle, 
\eneq
the full mean-field Hamiltonian can be written as
\beq \label{eq:Htot_SC} \mathcal{H}_{tot}=\sum_{\eta=\pm}\mathcal{H}^{\eta}_{BdG}+\sum_{\kk,\eta,e} [h_{-\eta}(-\kk)]_{ee} - \sum_{\kk \eta e_1 s_1 e_2 s_2} \langle \gamma^\dagger_{\kk e_1 \eta s_1 } \gamma^\dagger_{-\kk e_2 -\eta s_2}\rangle \Delta^{\eta}_{\kk; e_1 s_1 e_2 s_2 },
\eneq
where $\mathcal{H}^{\eta}_{BdG}$ is the Bogoliubov-de gennes (BdG) Hamiltonian of TBG. We only consider the spin singlet pairing channel, so 
\begin{eqnarray}
&\mathcal{H}^{\pm}_{BdG} =\frac{1}{2} \sum_{\kk\in MBZ} \psi_{\pm, \kk}^\dagger H_{BdG}^\pm(\kk) \psi_{\pm, \kk},\;\;\;\nonumber \\
& H_{BdG}^+(\kk) =\begin{pmatrix}
h_+(\kk)\otimes s_0 & 2\Delta_{\kk} \otimes (i s_y) \\
2\Delta_{\kk}^\dagger\otimes  (-i s_y) & - h_{-}^\star(-\kk) \otimes s_0
\end{pmatrix}\nonumber \\ &  H_{BDG}^-(\kk) =\begin{pmatrix}
h_-(\kk)\otimes s_0 & 2\Delta_{-\kk}^T \otimes (i s_y) \\
-2\Delta_{-\kk}^\star \otimes  (i s_y) & - h_{+}^\star(-\kk) \otimes s_0
\end{pmatrix}\nonumber 
\end{eqnarray} 
on the basis functions $\psi_{\eta, \kk}^\dagger = (\gamma^\dagger_{\kk, e_Y = \pm ,\eta,s =\uparrow \downarrow},\gamma_{-\kk, e_Y=\pm,-\eta,s=\uparrow \downarrow})$ at valley $\eta=\pm$ of TBG; $\gamma^\dagger_{\kk, e_Y ,\eta,s}$ is the electron creation operator for the flat bands in TBG with the Chern-band index $e_Y=\pm$ and spin $s$. 

Here we focus on the intra-Chern-band pairings so that the gap function is a $2\times 2$ diagonal matrix
\beq \Delta_{\kk} = \left(\begin{array}{cc} \Delta_{\kk,+}&0\\
0 & \Delta_{\kk,-}\end{array}\right), \eneq
where $\Delta_{\kk,e_Y}$ labels the gap function in the Chern-band channel $e_Y=\pm$. The intra-Chern-band pairing gap function $(\Delta_{\kk,+},\Delta_{\kk,-})$ belong to the 2D $E_2$ irrep and has the d-wave nature, which can be represented by \cite{liu2024electron}
\beq \label{eq:dwave_ansatz0} \Delta_{\kk,e_Y} = \Delta_{e_Y} \frac{k^2_{-e_Y}}{k^2+b^2} \eneq
with $k_{-e_Y}=k_x-i e_Y k_y$. In particular, $(\Delta_{+},\Delta_{-})= (\Delta_0,0)$ or $ (0,\Delta_0)$ gives the TR-breaking chiral d-wave while $(\Delta_{+},\Delta_{-})=(\Delta_0 , \Delta_0^* )$ represents the TR-preserving nematic Euler pairing. 

The superconducting particle-hole symmetry $\hat{\mathcal{P}}$ with $D(\hat{\mathcal{P}})=\rho_x \zeta_0 s_0$, requires
\beq -H_{BdG}^-(\kk) = \rho_x \zeta_0 s_0 H_{BdG}^{+\star}(-\kk) \rho_x \zeta_0 s_0, \eneq
where $s$, $\zeta$ and $\rho$ are the identity and Pauli matrices for the spin, Chern-band basis and superconducting particle-hole basis. 
One should note that the superconducting particle-hole operator $\hat{\mathcal{P}}$ is different from the unitary particle-hole symmetry operator $\hat{P}$
for the non-superconducting TBG Hamiltonian. As $H_{BdG}^-(\kk)$ can be related to $H_{BdG}^+(\kk)$ by particle-hole symmetry, we only focus on $H_{BdG}^+(\kk)$, which is an 8-by-8 matrix.  

$H_{BdG}^+(\kk)$ has a block diagonal form in the spin states, so we can further decompose it into two Hamiltonians, $H_{BdG}^+(\kk)=H_{BdG}^{+,+}(\kk)+H_{BdG}^{+,-}(\kk)$, where
\begin{eqnarray}\label{eq:HBdGpup}
& H_{BdG}^{+,\lambda=\pm}(\kk) = \frac{1}{2} \begin{pmatrix}
d_{0,\kk}-\mu & d_{x,\kk} & 2\lambda \Delta_{\kk,+} & 0 \\
d_{x,\kk} & d_{0,\kk}-\mu & 0 & 2\lambda \Delta_{\kk,-} \\
2\lambda \Delta^\star_{\kk,+} & 0 & - (d_{0,\kk}-\mu) & - d_{x,\kk} \\
0 & 2\lambda \Delta^\star_{\kk,-} & - d_{x,\kk} & - (d_{0,\kk}-\mu)
\end{pmatrix},
\end{eqnarray}
where $d_{0(x),\kk}=d_{0(x),+}(\kk)$. The eigen-energy $E^\pm_{\kk,n}$ of the BdG Hamiltonian $H_{BdG}^{+,+}(\kk)$ in (\ref{eq:HBdGpup}) can be solved analytically, 
\begin{eqnarray} \label{eq:BdGEnergy1}
& E^\pm_{\kk,n} = \pm \frac{1}{2}\sqrt{2A+\tilde{d}_{0,\kk}^2+d_{x,\kk}^2 + 2 n \sqrt{L_\kk }}
\end{eqnarray}
and
\beq L_\kk = A^2-4B^2+ \tilde{d}^2_{0,\kk}d^2_{x,\kk}+d_{x,\kk}^2(A-2B \cos(\Phi_{\kk})) ,
\eneq
where $n=\pm$, $\tilde{d}_{0,\kk}=d_{0,\kk}-\mu$, $A=|\Delta_{\kk,+}|^2+|\Delta_{\kk,-}|^2$, $B=|\Delta_{\kk,+}\Delta_{\kk,-}|$ and $\Phi_\kk=\phi_{\kk,-}-\phi_{\kk,+}$ with $\phi_{\kk,e_Y}$ the phase of order parameter $ \Delta_{\kk,e_Y}$, $\Delta_{\kk,e_Y}=|\Delta_{\kk,e_Y}|e^{i \phi_{\kk,e_Y}}$. 
For the gap function ansatz (\ref{eq:dwave_ansatz0}), we define $\Delta_{e_Y}=|\Delta_{e_Y}|e^{i\varphi_{e_Y}}$ and $\kk=k(\cos\theta_{\kk},\sin\theta_{\kk})$, and find 
\beq |\Delta_{\kk,e_Y}|=|\Delta_{e_Y}| \frac{k^2}{k^2+b^2}  \eneq
and
\beq \label{eq:phik_gapfunction} \phi_{\kk,e_Y}=\varphi_{e_Y}-2 e_Y \theta_{\kk}; \quad \Phi_{\kk}=\varphi_- - \varphi_+ + 4 \theta_{\kk}=\phi + 4\theta_\kk. \eneq
The momentum dependence in $\phi_{\kk,e_Y}$ represents the d-wave nature of the gap function and $\phi=\varphi_- - \varphi_+$ is the relative phase between $\Delta_-$ and $\Delta_+$. 


The above energy dispersion can have nodes when one of the following two scenarios is satisfied: (1)
\begin{eqnarray} 
&4B + \tilde{d}_{0,\kk}^2=d_{x,\kk}^2; \label{eq:nodecondi_11} \\
&|\Delta_{\kk,+}|=|\Delta_{\kk,-}|; \label{eq:nodecondi_12}\\
&\cos\left(\frac{\Phi_{\kk}}{2}\right)=0,
\label{eq:nodecondi_13}
\end{eqnarray}
or (2)
\begin{eqnarray} 
&4B = d_{x,\kk}^2; \label{eq:nodecondi_21} \\
&\tilde{d}_{0,\kk}=0; \label{eq:nodecondi_22} \\ 
&\cos\left(\frac{\Phi_{\kk}}{2}\right)=0. \label{eq:nodecondi_23} 
\end{eqnarray}

The case (1) corresponds to the condition of Euler pairing (Eq. (\ref{eq:nodecondi_12}))\cite{yu2022euler1,yu2022euler2}. 
As $\Phi_\kk$ relies on the phase $\phi_{\kk,e_Y}$ of $\Delta_{\kk,e_Y}$ (Eq.\ref{eq:phik_gapfunction}), which only depends on the momentum angle $\theta_\kk$ due to the d-wave nature in Eq. (\ref{eq:dwave_ansatz0}),  
the condition in Eq. (\ref{eq:nodecondi_13}) requires the nodes appearing at the momentum angle 
\beq \label{eq:nodeAngle1} \theta_\kk=\frac{1}{4} ((2n+1)\pi - \phi) \eneq
with $n\in \mathcal{Z}$ (any integer number). 
With Euler pairing condition in Eq. (\ref{eq:nodecondi_12}), $|\Delta_{\kk,+}|=|\Delta_{\kk,-}|=\tilde{\Delta}_{0,k}$, Eq. (\ref{eq:nodecondi_11}) is simplified to
\beq \label{Eq:Euler_constr1} 4\tilde{\Delta}_{0,k}^2 + \tilde{d}_{0,\kk}^2=d_{x,\kk}^2.  \eneq
At the momentum angles $\theta_\kk$ determined in Eq. (\ref{eq:nodeAngle1}), Eq. (\ref{Eq:Euler_constr1}) further fixes the momentum amplitude $|\kk|$ for the locations of the nodes. Therefore, the Euler pairing in Eq.(\ref{eq:nodecondi_12}) for the Hamiltonian (\ref{eq:HBdGpup}) can possess point nodes in the 2D momentum space with the locations determined by Eqs. (\ref{eq:nodeAngle1}) and (\ref{Eq:Euler_constr1}).  

For the case (2), Eq. (\ref{eq:nodecondi_23}) is the same as Eq. (\ref{eq:nodecondi_13}) and thus the momentum angle $\theta_\kk$ for the nodes, if exist, should also satisfy Eq. (\ref{eq:nodeAngle1}). At these momentum angles $\theta_\kk$, we still have two Eqs. (\ref{eq:nodecondi_21}) and (\ref{eq:nodecondi_22}), both of which depend on the momentum amplitude $|\kk|$ and thus cannot be generally satisfied at the same time. 
Therefore, we do not expect any node for the case (2). 
However, this scenario is relevant for a special case, namely the chiral limit $d_{0,\kk}=0$ and zero chemical potential $\mu=0$. In this case, $\tilde{d}_{0,\kk}=0$ in Eq. (\ref{eq:nodecondi_22}) is automatically satisfied for any momentum, and thus, Eq. (\ref{eq:nodecondi_21}) can fix the momentum amplitude $|\kk|$ at the momentum angles $\theta_\kk$ given in Eq. (\ref{eq:nodeAngle1}). In this case, it is possible that the nodes can appear for $|\Delta_{\kk,+}|\neq |\Delta_{\kk,-}|$. 

In the discussion below, we mainly consider the situation for non-zero $\mu$, and thus focus on the gapless condition of the case (1). 


The eigen-states of the BdG Hamiltonian can be encoded in the $U_\kk$ matrix defined by
\beq \label{eq:UMatrix_1}
U_\kk =\begin{pmatrix}
u_\kk & v_\kk \\
w_{\kk} & r_{\kk} 
\end{pmatrix}, \;\; H_{BdG}^{+,+}(\kk)= U_\kk D_\kk U^\dagger_\kk
\eneq
where $u_\kk, v_\kk, w_{\kk}, r_{\kk}$ are two-by-two matrices and $D_\kk$ is a diagonal matrix, $D_\kk=\text{Diag}(E^+_+,E^+_-,E^-_+,E^-_-)$, with $E^+_{n}>0$ and $E^-_{n}<0$ ($n=\pm$). 
The Bogoliubov quasi-particle operators are defined by
\beq \label{eq:Bogoliubov_transformation_1}
\left(\begin{array}{c} \gamma_{\kk, e_1 ,+,\uparrow } \\ \gamma^\dagger_{-\kk, e_1, - ,\downarrow} \end{array}\right)= \sum_{n} [U_\kk]_{e_1 n} \left(\begin{array}{c} \alpha_{\kk, n} \\ \beta^\dagger_{-\kk, n} \end{array}\right)
\eneq 
so that
\begin{eqnarray}\label{eq:HBdG_valleyp_lambdap}
& \mathcal{H}_{BdG}^{+, +}= \sum_{\kk, e_1 e_2} (\alpha^\dagger_{\kk, e_1 } \; \;\beta_{-\kk, e_1}) [D_\kk]_{e_1,e_2}\left(\begin{array}{c} \alpha_{\kk, e_2  } \\ \beta^\dagger_{-\kk, e_2} \end{array}\right) \nonumber \\
& = \sum_{\kk, n} \left( E^+_{\kk,n} \alpha^\dagger_{\kk, n}  \alpha_{\kk, n }+ (- E^-_{\kk, n}) \beta^\dagger_{-\kk, n} \beta_{-\kk, n}  \right)
+ \sum_{\kk,n } E^-_{\kk,n}. 
\end{eqnarray}
The superconductor (SC) ground state $|GS\rangle$ is defined as $\alpha_{\kk,n}\ket{GS} = \beta_{\kk,n}\ket{GS} =0$
with $n= \pm$, and $\alpha^\dagger_{\kk,n}$ and $\beta^\dagger_{\kk,n}$ create Bogoliubov quasi-particles with the energy $E^+_{\kk,n}$ and $-E^-_{\kk,n}$, respectively. The SC condensation energy is then defined as 
\beq E_c=\langle \mathcal{H}_{tot} \rangle_s-\langle \mathcal{H}_{tot} \rangle_n, \eneq
where $\langle \mathcal{H}_{tot} \rangle_s=\langle GS|\mathcal{H}_{tot}|GS \rangle$ is the superconductor ground state energy, and
$\langle \mathcal{H}_{tot} \rangle_n$ is the energy of the normal metallic state, which can be obtained from $\langle \mathcal{H}_{tot} \rangle_s$ by 
taking the superconductor gap function $\Delta$ to be zero.

The condensation energy (per particle) for the intra-Chern-band channels can be simplified as \cite{liu2024electron}
\beq \label{eq:condensation_energy_1} E_c= \frac{4}{N_M} \left(\sum_{\kk, n} (E^-_{\kk,n}-\xi^-_{\kk,n})  +  \sum_{\kk e_Y e'_Y n }  u_{\kk,e_Y n}^* w_{\kk, e'_Y n}  \Delta_{\kk; e_Y e'_Y } \right),  \eneq
where $\xi^-_{\kk,n}=-\frac{1}{2}\left|\tilde{d}_{0,\kk}+n d_{x,\kk}\right|$ ($n=\pm$) is the single-particle eigen-energy without superconductivity. 

To determine the exact pairing form for the temperature well below $T_c$, we also derive the full self-consistent gap equation \cite{liu2024electron} for the intra-Chern-band channel as 
\beq \label{eq:gapequation_G1} \Delta_{\kk;e_Y}= \frac{1}{N_M} \sum_{\kk', n } V_{\kk\kk'}^{+ e_Y e_Y}  u_{-\kk',e_Y n} w^*_{-\kk',e_Y n} \eneq
at zero temperature, where $V_{\kk\kk'}^{\eta=\pm, e_Y e'_Y}$ is the attractive electron-electron interaction mediated by $K$-phonons. 
From the heavy fermion model \cite{song2022magic,shi2022heavy}, it has been shown that the $V$-interaction for the intra-Chern-band channels has the form \cite{liu2024electron}
\begin{eqnarray}\label{eq:Interactionform_intra_Chern1}
V^{\eta,e_Y,e_Y}_{\kk,\kk'}=U^*_{1,e_Y,\kk} U_{1,e_Y,\kk'}; \; U_{1,e_Y,\kk}=\frac{\sqrt{V_0}}{k^2+b^2} k_{e_Y}^2;  
\end{eqnarray}
where $k_{e_Y}=k_x+i e_Y k_y=k e^{i e_Y \theta_\kk}$, $e_Y=\pm$, and $V_0$ and $b$ are material dependent parameters with their values $V_0=0.4 meV$ and $b=0.185 k_\theta$ \cite{liu2024electron}. 
Here $k_\theta=2|{\bf K}_D|\sin\frac{\theta}{2}$ and $\theta=1.06^\circ$ is the twist angle between two layers and $\KK_D=\frac{4\pi}{3a_0}(1,0)$ with $a_0$ the lattice constant of graphene. The above form of interaction can lead to the d-wave ansatz in Eq. (\ref{eq:dwave_ansatz0}).
With the above form of $V$-interaction and gap function ansatz (\ref{eq:dwave_ansatz0}), the gap equation (Eq. \ref{eq:gapequation_G1}) can be simplified to 
\beq \label{eq:GapEquation_intra_dwave1} \Delta_{e_Y}= \frac{V_0 }{N_M} \sum_{\kk', n } \frac{{k'}^2_{e_Y}}{k'^2+b^2}  u_{-\kk',e_Y n} w^*_{-\kk',e_Y n}. \eneq
Using Eq. (\ref{eq:dwave_ansatz0}) and (\ref{eq:GapEquation_intra_dwave1}), we find the condensation energy in Eq. (\ref{eq:condensation_energy_1}) can be simplified as \cite{liu2024electron}
\beq \label{eq:condensation_energy_intra1} 
E_c=4 \left( \frac{1}{N_M} \sum_{\kk, n} (E^-_{\kk,n}-\xi^-_{\kk,n})  + \frac{1}{V_0} \sum_{e_Y } |\Delta_{e_Y}|^2 \right).  \eneq

\section{Numerical Results for self-consistent gap equation} \label{Sec:Numerical_gapequation}
\begin{figure}
	\includegraphics[width=0.9\textwidth,angle=0]{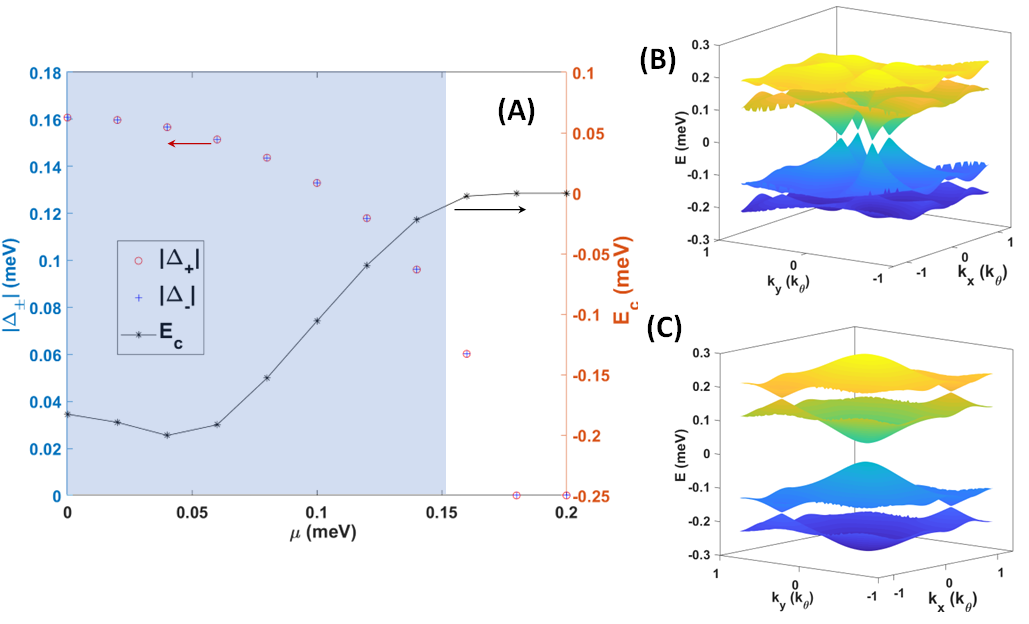}
	\centering
	\caption{ The self-consistent solution of the gap equation for the single-particle bandwidth parameter $\xi=0.6$. (A) Red circles and blue crosses show the gap functions $\Delta_{+}$ and $\Delta_{-}$ of the self-consistent gap equation as a function of the chemical potential $\mu$ for the intra-Chern-band pairing. The BdG spectrum $E^{\pm}_{\kk,n=1,2}$ has nodes in the shaded region. Black stars show the condensation energy $E_c$ as a function of $\mu$. The self-consistent solution is always Euler pairing once the SC order parameter is non-zero. (B) and (C) show the BdG spectrum $E^{\pm}_{\kk,n=1,2}$ for the chemical potential $\mu=0.04 meV$ and $0.16 meV$, respectively. }
	\label{fig_gapself_3}
\end{figure}

\begin{figure}
	\includegraphics[width=0.9\textwidth,angle=0]{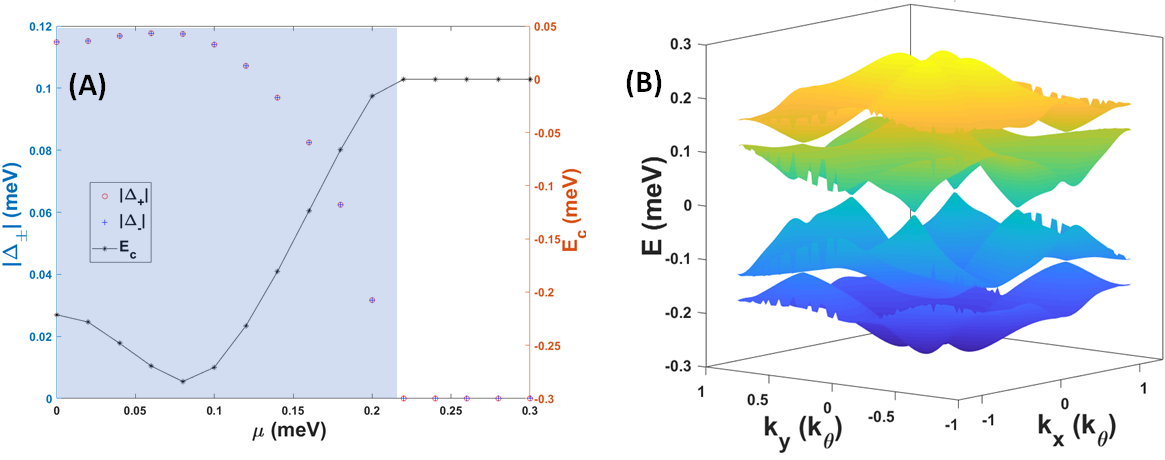}
	\centering
	\caption{ The self-consistent solution of the gap equation for the single-particle bandwidth parameter $\xi=1$. (A) Red circles and blue crosses show the gap functions $\Delta_{+}$ and $\Delta_{-}$ of the self-consistent gap equation as a function of the chemical potential $\mu$ for the intra-Chern-band pairing. The BdG spectrum has nodes in the shading regime. Black stars show the condensation energy $E_c$ as a function of $\mu$.  (B) shows the BdG spectrum for the chemical potential $\mu=0.08 meV$.   }
	\label{fig_gapself_2}
\end{figure}

In this section, we will present our numerical results for the full gap equation (Eq. \ref{eq:gapequation_G1}) for the intra-Chern-band channel. Our numerical results are performed for the chiral limit ($d_{0,\eta}(\kk)=0$), so that the inter-Chern-band and intra-Chern-band channels are decoupled and can be studied separately. For a non-zero chemical potential $\mu$, the gapless condition (\ref{eq:nodecondi_22}) for the case (2) cannot be satisfied, and thus we below focus on the case (1) for the Euler pairing.  
We emphasize that the motivation of this work is {\it not} to simulate the realistic TBG systems, but instead to understand the stability of possible pairing states for the flat bands in a theoretical model. As the SC property sensitively depends on the band width of the flat bands, we introduce a parameter $\xi$ to manually tune the bandwidth of the single particle Hamiltonian. Thus, the BdG Hamiltonian is written as
\begin{eqnarray}\label{eq:HBdGpup_1}
& H_{BdG}^{+,\lambda=\pm}(\kk) = \frac{1}{2} \begin{pmatrix}
-\mu & \xi d_{x,\kk} & 2\lambda \Delta_{\kk,+} & 0 \\
\xi d_{x,\kk} & -\mu & 0 & 2\lambda \Delta_{\kk,-} \\
2\lambda \Delta^\star_{\kk,+} & 0 & \mu & - \xi d_{x,\kk} \\
0 & 2\lambda \Delta^\star_{\kk,-} & - \xi d_{x,\kk} & \mu
\end{pmatrix},
\end{eqnarray}
where $d_{0,\kk}=0$ in the chiral limit. This BdG Hamiltonian (\ref{eq:HBdGpup_1}) can be solved numerically, to obtain the eigen-states for the $U_\kk$ matrix in Eq. (\ref{eq:UMatrix_1}). With the form of $U_\kk$ matrix, we can use the gap equation (\ref{eq:GapEquation_intra_dwave1}) to evaluate the gap function $\Delta_{e_Y}$, which can be substituted back into the BdG Hamiltonian (\ref{eq:HBdGpup_1}), together with the d-wave ansatz in Eq. (\ref{eq:dwave_ansatz0}), to update the eigen-state matrix $U_\kk$. This process allows us to obtain a self-consistent solution for the gap functions. Fig. \ref{fig_gapself_3}(A) shows the self-consistent solutions of the gap functions and the corresponding condensation energies calculated from Eq. (\ref{eq:condensation_energy_intra1}) for $\xi=0.6$ (corresponding to unrealistically small bandwidth around $0.3 meV$) at zero temperature. As $|\Delta_{\kk,+}|=|\Delta_{\kk,-}|$ is always satisfied in Fig. \ref{fig_gapself_3}(A), the nematic Euler SC phase is the self-consistent solution in a wide chemical potential range when SC order parameter is non-zero. The point nodes can exist in the shadowed regime of the chemical potentials, corresponding to the gapless condition of the case (1). The BdG spectrum for the chemical potential $\mu=0.04$ meV is shown in Fig. \ref{fig_gapself_3}(B), where four point nodes can be found. With increasing the chemical potential $\mu$, the nodes can move in the momentum space. When the chemical potential $\mu$ is larger than $0.15$meV (outside the shadowed regime), four nodes annihilate and the BdG spectrum become fully gapped, as shown in Fig. \ref{fig_gapself_3}(C) for $\mu=0.16$ meV. With further increasing the chemical potential $\mu$, the SC order parameter drops to zero around $\mu\approx 0.18$meV and the system becomes non-superconducting. The temperature dependence of the gap function and condensation energy can be found in Fig. 1 of the main text. Fig. \ref{fig_gapself_2}(A) shows the gap function and the corresponding condensation energy for $\xi=1$ (corresponding to bandwidth around $0.5 meV$) and we also find the nematic Euler SC phase is energetically favored. In this situation, we find the fully gapped SC regime disappears, so our model is in the nodal SC phase for $\mu<0.22$meV and in the non-superconducting phase for $\mu>0.22$meV.  A typical BdG spectrum is shown in Fig. \ref{fig_gapself_2} (B) for $\mu=0.08$ meV. 



As the self-consistent solution reveals equal pairing amplitude for $\Delta_{\kk,+}$ and $\Delta_{\kk,-}$, occurring at non-zero $\mu$, the nodal SC phase should correspond to the case (1). Below we will numerically test how the condensation energy $E_c$ depends on the form of the gap function $\Delta_{\kk,e_Y}$ and then provide the theoretical analysis of the stability of Euler pairing in Sec. \ref{Sec:CondenstationEnergy}. 
We consider the following general form of the gap function ansatz in Eq. (11) in the main text, or equivalently
\beq \label{eq:gap_function_expansion} \Delta_{\kk,+}=\Delta_{0}\cos\alpha e^{-i\phi/2} \frac{k^2_{-}}{k^2+b^2} = \Delta_{0\kk}\cos\alpha e^{- i(\phi/2+ 2 \theta_\kk)}, \; \;  \Delta_{\kk,-}=\Delta_{0}\sin\alpha e^{i\phi/2} \frac{k^2_{+}}{k^2+b^2} = \Delta_{0\kk}\sin \alpha e^{i (\phi/2 + 2 \theta_\kk)}, \eneq 
where $\Delta_{0\kk}=\Delta_0 \frac{k^2}{k^2+b^2}$ with $\Delta_{0}$ to be a real number, $\theta_\kk$ is the angle of the momentum $\kk$, $\alpha$ is a parameter that tunes the amplitude of two components for the 2D $E_2$ irrep gap function, and $\phi$ is the relative phase factor between $\Delta_-$ and $\Delta_+$. 
The Euler pairing occurs at $\alpha=\pi/4$ while the chiral d-wave pairing occurs at $\alpha=0$ or $\pi/2$. Other values of $\alpha$ gives more general form for 2D $E_2$ irrep gap function. Fig. \ref{fig_condeng_comp_3}(A) shows the condensation energy $E_c$ as a function of $\Delta_0$ and $\alpha$ for $\mu=0.04$ meV, $\xi=0.6$ and $\phi=0$. Our numerical results show the minimal value of $E_c$ occurring at $\Delta_0 \approx 0.22$ meV, coinciding well with the value from our self-consistent equations in Fig. \ref{fig_gapself_3}(A) ($\Delta_0=\sqrt{2} |\Delta_+|$). We further plot $E_c$ as a function of $\alpha$ and $\phi$ for $\Delta_0 \approx 0.22$ meV in Fig. \ref{fig_condeng_comp_3}(B), from which the condensation energy $E_c$ has a strong dependence on $\alpha$ with its minimum occurring at $\alpha=\pi/4$, but a very weak dependence on the relative phase $\phi$. We find the variation of $E_c$ with respect to $\phi$ is only around 1\% of the minimal value of $|E_c|\sim 0.2$ meV. This weak $\phi$ dependence can be understood as follows. From Eqs. (\ref{eq:condensation_energy_intra1}) and (\ref{eq:BdGEnergy1}), one can see that the $\phi$ dependence of $E_c$ is via the momentum angle dependence in $E^-_{\kk,n}$. Particularly, one can consider the rotation of momentum coordinate by an angle $\phi/4$, 
\begin{eqnarray}\label{eq:coordinateTransformation1}
&&\begin{pmatrix}
k_x' \\
k_y' 
\end{pmatrix}
=\begin{pmatrix}
\cos(\phi/4) & -\sin(\phi/4) \\
\sin(\phi/4) & \cos(\phi/4)
\end{pmatrix} 
\begin{pmatrix}
k_x \\
k_y 
\end{pmatrix}, 
\end{eqnarray}
which transforms the momentum angle by $\theta_{\kk'}=\theta_\kk+\phi/4$ but preserves momentum amplitude $k'=k$. Thus in the new momentum coordinate, $\Phi_{\kk'}=4\theta_{\kk'}$ which means the $\phi$ dependence in $\Phi_{\kk'}$ of $E^\pm_{\kk',n}$ (Eq.\ref{eq:BdGEnergy1}) is removed. If we take the isotropic approximation for the energy spectrum of the flat bands in the chiral limit, we have $d_{0,\kk}=0$ and $d_{x,\kk}\approx C_0-C_2 k^2$ around $\Gamma_M$, and then in the new coordinate, $d_{x,\kk'}=C_0-C_2 k'^2$ is independent of $\phi$. Thus, $E_c$ in Eq.(\ref{eq:condensation_energy_intra1}) does not depend on $\phi$ after transforming the summation over $\kk$ to that over $\kk'$. From this discussion, we conclude that the $\phi$ dependence of $E_c$ originates from the in-plane anisotropic energy spectrum of the flat bands, which is weak when the Fermi energy is close to the band top of the flat bands around $\Gamma_M$.

Fig. \ref{fig_condeng_comp_3}(C) and (D) show the energy dispersion of the BdG spectrum $E^{\pm}_{\kk,n=1,2}$ for the chiral d-wave pairing $\alpha=0$ and the Euler pairing $\alpha=\pi/4$, respectively, for comparison. Although the Euler pairing is gapless around point nodes, its BdG gap is much larger when the momenta are away from these point nodes, as compared with the gap of the chiral d-wave pairing. Thus, to understand the stability of the Euler pairing, we need to analyze the gap function over the whole Moir\'e Brillouin zone (MBZ), as discussed in the next Sec. \ref{Sec:CondenstationEnergy}. As the condensation energy is independent of $\phi$ in  \ref{fig_condeng_comp_3}(B), we below always choose $\phi=0$.


\section{Analysis of Condensation Energy around Gapless Nematic Phase}\label{Sec:CondenstationEnergy}
Now we will perform a theoretical analysis for the condensation energy $E_c$ (Eq. \ref{eq:condensation_energy_intra1}) around the gapless nematic phase that we have found for the case (1) and analyze the stability of the gapless Euler pairing phase, to understand our numerical results in Sec. \ref{Sec:Numerical_gapequation}. 
We consider gap function form in Eq. (\ref{eq:gap_function_expansion}), with the value of $\Delta_{0}$ either to be solved from the self-consistent equations numerically, as discussed in Sec. \ref{Sec:Numerical_gapequation}, or equivalently, by minimizing the condensation energy $E_c$ with respect to $\Delta_0$ in Eq. (\ref{eq:condensation_energy_intra1}). As both numerical results are consistent and suggest the nematic Euler pairing is a stable solution in a wide range of chemical potentials, we will expand the gap function around the Euler pairing by choosing $\alpha=\frac{\pi}{4}+\delta_0$, where $\delta_0\ll 1$, and demonstrate the Euler pairing with $\alpha=\pi/4$ is at least a local minimum of condensation energy $E_c$. Our task is to evaluate the condensation energy $E_c$ in Eq. (\ref{eq:condensation_energy_intra1}) perturbatively in orders of $\delta_0$, namely 
\beq \label{eq:Ec_expansion} 
E_c=E_{c0}+\gamma \delta_0 +\beta \delta^2_0,  \eneq
up to the $\delta_0^2$ order and determine the expressions for $\gamma$ and $\beta$. The $\xi^-_{\kk,n}$ term in $E_c$ does not depend on $\Delta_{\kk,e_Y}$, so it only contributes a constant term to $E_{c0}$. The last term in Eq. (\ref{eq:condensation_energy_intra1}) can be easily evaluated as
\beq \frac{N_M}{V_0} \sum_{e_Y } |\Delta_{e_Y}|^2 = \frac{N_M}{V_0} \Delta^2_0, \label{eq:Energy_MF1}
\eneq
which is independent of $\delta_0$ and only contributes to $E_{c0}$. 

\begin{figure}
\includegraphics[width=0.9\textwidth,angle=0]{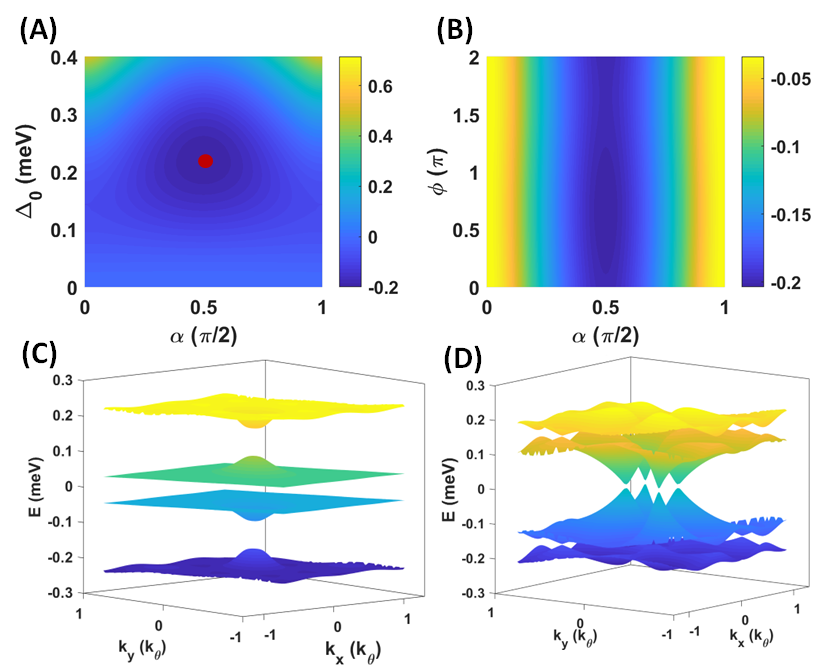}
	\centering
	\caption{ (A) Condensation energy $E_c$ as a function of $\alpha$ and $\Delta_0$ for $\phi=0$. Red dot depicts the position of the minimal value for $E_c$. (B) Condensation energy $E_c$ as a function of $\alpha$ and $\phi$ for $\Delta_0=0.22$ meV, which is the gap value from self-consistent equations. Here $(\Delta_+,\Delta_-)=\Delta_0(\cos\alpha e^{-i\phi/2},\sin\alpha e^{i\phi/2})$.  (C) and (D) show the energy spectrum of the BdG Hamiltonian for chiral d-wave pairing $\alpha=0$ and Euler pairing $\alpha=0.5$ with $\phi=0$. Here $\mu=0.04 meV$ and $\xi=0.6$.       }
	\label{fig_condeng_comp_3}
\end{figure}

Our main task below is to perform the perturbation expansion of the eigen-energy $E^-_{\kk,n}$ in Eq. (\ref{eq:BdGEnergy1}) to evaluate the parameters $\gamma$ and $\beta$ in the expansion (\ref{eq:Ec_expansion}) of $E_c$. We will show $\gamma=0$ and $\beta>0$ below. 

From 
\beq A=\sum_{e_Y} |\Delta_{\kk,e_Y}|^2=\Delta_{0\kk}^2, \;\; B\approx \frac{\Delta_{0\kk}^2}{2}(1-2\delta_0^2), 
\eneq
we find 
\begin{eqnarray}
&& L_\kk = A^2-4B^2+ \tilde{d}^2_{0,\kk}d^2_{x,\kk}+d_{x,\kk}^2(A-2B \cos(\Phi_{\kk})) \nonumber\\
&& \approx d^2_{x,\kk}(\tilde{d}_{0,\kk}^2+\Delta_{0\kk}^2(1-\cos(\Phi_{\kk})))+(4\Delta^2_{0\kk}+2d_{x,\kk}^2\cos(\Phi_{\kk}))\Delta^2_{0\kk}\delta_0^2. 
\end{eqnarray} 
The above expression has been arranged in orders of $\delta_0$ and we take both $\tilde{d}_{0,\kk}$ and $d_{x,\kk}$ are non-zero, which can be possible for non-zero $\mu$, so that we can perform the perturbation expansion. It should be noted that $d_{x,\kk}$ is zero for $\kk=\KK_M$ in the TBG model but we can consider the regime when the Fermi energy is around $\Gamma_M$ point in the Moir\'e BZ, so that the relevant momentum regime is away from $\KK_M$. With the above consideration, we expand $\sqrt{L_\kk}$ as
\begin{eqnarray} 
&& \sqrt{L_\kk} \approx |d_{x,\kk}|\sqrt{\tilde{d}_{0,\kk}^2+2\Delta_{0\kk}^2\sin^2(\Phi_\kk/2)} \left[ 1 +\frac{\Delta_{0\kk}^2(2\Delta_{0\kk}^2+d_{x,\kk}^2\cos(\Phi_\kk))}{d_{x,\kk}^2(\tilde{d}_{0,\kk}^2+2\Delta_{0\kk}^2\sin^2(\Phi_\kk/2))}\delta_0^2\right]
\end{eqnarray}
and the corresponding negative eigen-energy $E^-_{\kk,n}$ as
\beq E^-_{\kk,n}\approx -\frac{1}{2}\sqrt{2\Delta_{0\kk}^2+\tilde{d}_{0,\kk}^2+d_{x,\kk}^2+2n \sqrt{L_\kk}}
\eneq
up to the order of $\delta_0^2$ ($n=\pm$). For $\delta_0=0$, the energy dispersion becomes
\begin{eqnarray}
&& E^-_{\kk,n} = -\frac{1}{2}\sqrt{2\Delta_{0\kk}^2+\tilde{d}_{0,\kk}^2+d_{x,\kk}^2+2n |d_{x,\kk}|\sqrt{\tilde{d}_{0,\kk}^2+2\Delta_{0\kk}^2\sin^2(\Phi_\kk/2)}}\nonumber\\
&& = -\frac{1}{2}\sqrt{ \left(|d_{x,\kk}|+n\sqrt{\tilde{d}_{0,\kk}^2+2\Delta_{0\kk}^2\sin^2(\Phi_\kk/2)}\right)^2 + 2\Delta_{0\kk}^2\cos^2(\Phi_\kk/2)}  . 
\end{eqnarray} 
The energy spectrum of $E^-_{\kk,n}$ has two branches $n=\pm$, and the $n=-$ branch has points nodes at the momentum location $\kk_g$ determined by
\beq \label{eq:gapless_condition} \cos(\Phi_{\kk_g}/2)=0, \; \;  d_{x,\kk_g}^2=\tilde{d}_{0,\kk_g}^2+2\Delta_{0\kk_g}^2. \eneq
This is nothing but the node conditions (\ref{eq:nodecondi_11}) and (\ref{eq:nodecondi_13}) for the case (1). The case (2) cannot be satisfied for non-zero chemical potential $\mu$. 

To obtain $E_c$ for non-zero $\delta_0$, we need to evaluate $\sum_{n,\kk}E^-_{\kk,n}$. For the momentum summation, we separate the MBZ into two regions. One is for the region around the point node $\kk_g$, namely $\kk=\kk_g+\qq$ with $|\qq|<\varepsilon$ where $\varepsilon$ is a small number serving as the momentum cut-off, and this momentum region is denoted as $\Omega_{\kk_g}$. The other is the momentum region away from the point nodes, namely $\kk \not\in \Omega_{\kk_g}$. 
The momentum summation for these two regions need to be evaluated separately. 
In the region $\kk \not\in \Omega_{\kk_g}$, the BdG spectrum always has a gap, so we can perform the perturbation expansion in $\delta_0$ for $E^-_{\kk,n}$. We denote 
\beq \label{eq:f0_eigenenergy}
f_{0\kk,n}=\frac{1}{2} \sqrt{ \left(|d_{x,\kk}|+n\sqrt{\tilde{d}_{0,\kk}^2+2\Delta_{0\kk}^2\sin^2(\Phi_\kk/2)}\right)^2 + 2\Delta_{0\kk}^2\cos^2(\Phi_\kk/2) },
\eneq
and
\beq \label{eq:fun2kn} f_{2\kk,n}= \frac{n}{2} \frac{\Delta_{0\kk}^2(2\Delta_{0\kk}^2+d_{x,\kk}^2\cos(\Phi_\kk))}{|d_{x,\kk}|\sqrt{\tilde{d}^2_{0,\kk}+2\Delta_{0\kk}^2\sin^2(\Phi_\kk/2)}}, \eneq
and can expand $E^-_{\kk,n}$ in terms of $\delta_0$ in the momentum region $\kk \not\in \Omega_{\kk_g}$ as
\begin{eqnarray}\label{eq:Eng_expansion}
E^-_{\kk,n} \approx -\sqrt{f^2_{0\kk,n}+f_{2\kk,n}\delta_0^2}\approx -f_{0\kk,n}-\frac{f_{2\kk,n}}{2f_{0\kk,n}}\delta_{0}^2 
\end{eqnarray}
up to the order of $\delta_0^2$ for $n=\pm$. In this region ($\kk \not\in \Omega_{\kk_g}$), only $\delta_0^2$ term appears in $E^-_{\kk,n}$. 

For the momentum region around the point nodes, $\kk \in \Omega_{\kk_g}$, the $n=+$ branch is fully gapped, so this branch can be expanded perturbatively in the same form as Eq. (\ref{eq:Eng_expansion}). 
Thus, for $\kk \in \Omega_{\kk_g}$, we have \begin{eqnarray}
E^+_{\kk\in \Omega_{\kk_g},+} \approx - f_{0\kk,+} - \frac{f_{2\kk,+}}{2f_{0\kk,+}}\delta_{0}^2 
\end{eqnarray}
up to the order of $\delta_0^2$, as $f_{0\kk,+}\neq 0$.

Next we focus on the $n=-$ branch for $\kk \in \Omega_{\kk_g}$.  With Eqs. (\ref{eq:f0_eigenenergy}) and (\ref{eq:fun2kn}), we have at $\kk=\kk_g$  
\begin{eqnarray}
&&f_{0\kk_g,-}=0 \\
&&f_{2\kk_g,-}= - \frac{1}{2} \frac{\Delta_{0\kk_g}^2(2\Delta_{0\kk_g}^2-d_{x,\kk_g}^2)}{|d_{x,\kk_g}|\sqrt{\tilde{d}^2_{0,\kk_g}+2\Delta_{0\kk_g}^2}} =\frac{\Delta_{0\kk_g}^2\tilde{d}_{0,\kk_g}^2}{2d_{x,\kk_g}^2}.
\end{eqnarray}
Due to the vanishing value of $f_{0\kk_g,-}$, we cannot perform the perturbation expansion in this region. Exactly at $\kk_g$, we find 
\beq E^-_{\kk_g,-} \approx -\sqrt{f^2_{0\kk_g,-}+f_{2\kk_g,-}\delta_{0}^2} = -  \left|\frac{\Delta_{0\kk_g}\tilde{d}_{0,\kk_g}}{\sqrt{2}d_{x,\kk_g}^2} \delta_0\right|. \label{eq:EigenEng_gap}
\eneq
One can see that $\delta_0$ term in the gap function can induce a gap opening that is proportional to $\delta_0$ and thus lower the energy at $\kk_g$. At the first sight, it seems that this energy reduction is proportional to $\delta_0$. However, as we will show below, this is {\it not} true because the condensation energy $E_c$ involves the momentum integral over MBZ for $E^-_{\kk,-}$. Thus, we need to consider the energy $E^-_{\kk,-}$ in the whole region $\Omega_{\kk_g}$ around $\kk_g$. We take $\kk=\kk_g+\qq$ with $\qq$ a small number. Moreover, $\delta_{0}$ is also a small number and thus for the lowest order in both $\delta_{0}$ and $\qq$, we keep $\kk$ to be $\kk_g$ for $f_{2\kk,-}$, but expand $f_{0\kk=\kk_g+\qq,-}$ for a small $\qq$. From our numerical calculations in Fig. \ref{fig_condeng_comp_3}B, we find the condensation energy $E_c$ has a very weak dependence on the relative phase $\phi$. Thus, our strategy is to evaluate the expansion of $E_c$ in Eq.(\ref{eq:Ec_expansion}) and extract the parameters $E_{c0}$, $\beta$ and $\gamma$ for a fixed $\phi$. Without loss of generality, we will first choose $\phi=0$ and then discuss other $\phi$ values. With $\phi=0$ in Eq. (\ref{eq:phik_gapfunction}),  $\Phi_{\kk}=4\theta_\kk$ with $\theta_\kk$ the angle of the momentum $\kk$, and the momentum angles for the nodes from Eq. (\ref{eq:nodeAngle1}) become
$\theta_{\kk_g} = \frac{1}{4}(2n+1)\pi$ with integer $n$ or equivalently
\beq \label{eq:thetakgvalue}\theta_{\kk_g}=\pm \frac{\pi}{4}, \pm \frac{3\pi}{4}. \eneq
We also denote $|\kk_g|=k_0$ with the value of $k_0$ determined by the second condition of Eq. (\ref{eq:gapless_condition}) or Eq. (\ref{eq:nodecondi_11}). The four point nodes are located at $\kk_{g1}=\frac{k_0}{\sqrt{2}}(1,1)$, $\kk_{g2}=\frac{k_0}{\sqrt{2}}(1,-1)$, $\kk_{g3}=\frac{k_0}{\sqrt{2}}(-1,1)$ and $\kk_{g4}=\frac{k_0}{\sqrt{2}}(-1,-1)$ according to the values of $\theta_{\kk_g}$ in Eq. (\ref{eq:thetakgvalue}).

We need to expand $E^-_{\kk,-}$ around $\kk_{gi}$ ($i=1,2,3,4$) up to the lowest order in $\qq$. Here we take $\kk_{g1}$ as an example and the expansion around other node momenta is similar. Around $\kk_{g1}$ and up to linear order in $\qq$, we have the following expansions
\begin{eqnarray} 
&& \Delta_{0\kk}\cos(\Phi_{\kk}/2)=\Delta_{0\kk}\cos(2\theta_{\kk})=\Delta_0 \frac{k^2}{k^2+b^2}(\cos^2\theta_\kk-\sin^2\theta_\kk) =\Delta_0 \frac{k^2}{k^2+b^2}((k_x/k)^2-(k_y/k)^2) = \frac{\Delta_0}{k^2+b^2}(k^2_x-k^2_y)  \nonumber \\
&& = \frac{\Delta_0}{(k_{g1,x}+q_x)^2+(k_{g1,y}+q_y)^2+b^2}((k_{g1,x}+q_x)^2-(k_{g1,y}+q_y)^2)  \approx \frac{\Delta_0 \sqrt{2} k_0 (q_x - q_y) }{k_0^2+\sqrt{2} k_0 (q_x+q_y) +b^2} \nonumber \\
&& \approx \frac{\sqrt{2}\Delta_0k_0}{k_0^2+b^2} (q_x-q_y)= \frac{\sqrt{2} \Delta_{0\kk_g}}{k_0} (q_x-q_y), 
\end{eqnarray}
and
\begin{eqnarray} && \Delta_{0\kk}\sin(\Phi_{\kk}/2)=\Delta_{0\kk}\sin(2\theta_{\kk}) = 2\Delta_0 \frac{k_x k_y}{k^2+b^2} \approx 2\Delta_0 \frac{(k_{g1,x}k_{g1,y}+k_{g1,x}q_x+k_{g1,y}q_y)}{|\kk_{g1}|^2+2(k_{g1,x}q_x+k_{g1,y}q_y)+b^2} \nonumber \\
&& \approx \Delta_0 \left( \frac{k_0^2}{k_0^2+b^2} + \frac{\sqrt{2}b^2 k_0}{(k_0^2+b^2)^2} (q_x+q_y) \right) = \Delta_{0\kk_g} \left( 1 + \frac{\sqrt{2}b^2 }{k_0(k_0^2+b^2)} (q_x+q_y) \right).
\end{eqnarray}

In the chiral limit, $d_{0,\kk}=0$ and $\tilde{d}_{0,\kk}=-\mu$, and we assume the chemical potential is close to the top of the bands located at $\Gamma_M$ in Moir\'e BZ, so we can approximate $d_{x,\kk}\approx C_0-C_2 k^2$, which leads to
\beq d_{x,\kk}\approx d_{x,\kk_{g1}} + (\partial d_{x,\kk}/\partial \kk)_{\kk_{g1}}\cdot \qq = d_{x,\kk_{g1}} - 2 C_2 \kk_{g1}\cdot \qq = d_{x,\kk_{g1}} - \sqrt{2} C_2 k_0 (q_x+q_y). 
\eneq
With the above expansions, we now can find 
\beq \sqrt{\tilde{d}_{0,\kk}^2+2\Delta_{0\kk}^2\sin^2(\Phi_{\kk}/2)} \approx |d_{x,\kk_{g1}}| \left(1 + \frac{2\Delta_{0\kk_{g}}^2}{d_{x,\kk_{g1}}^2} \frac{\sqrt{2}b^2}{k_0(k_0^2+b^2)} (q_x+q_y) \right) \eneq
\begin{eqnarray} 
&& f_{0\kk,-}^2=\frac{1}{4} \left[ \left(|d_{x,\kk}|-\sqrt{\tilde{d}_{0,\kk}^2+2\Delta_{0\kk}^2\sin^2(\Phi_{\kk}/2)}\right)^2 + 2\Delta_{0\kk}^2\cos^2(\Phi_{\kk}/2) \right]\nonumber\\
&& \approx \frac{1}{2} \left( - C_2 k_0 -  \frac{2\Delta_{0\kk_g}^2}{|d_{x,\kk_{g1}}|} \frac{b^2}{k_0(k_0^2+b^2)}  \right)^2 (q_x+q_y)^2 + \frac{ \Delta^2_{0\kk_g}}{k^2_0} (q_x-q_y)^2 . 
\end{eqnarray}
Now let us define 
\beq q_1=(q_x+q_y)/\sqrt{2}, \; q_2=(q_x-q_y)/\sqrt{2}, \; v_1=C_2k_0+ \frac{2\Delta_{0\kk_g}^2}{|d_{x,\kk_{g1}}|} \frac{b^2}{k_0(k_0^2+b^2)} , 
\; v_2=\frac{ \sqrt{2}\Delta_{0\kk_g}}{k_0},  B_3=\left|\frac{\Delta_{0\kk_g}\mu}{\sqrt{2}d_{x,\kk_{g1}}}\right|
\eneq 
and
we obtain the expansion for $E^-_{\kk,-}$ around $\kk_g$ as
\begin{eqnarray} \label{eq:eigen_energy_expansion1}
&& E^-_{\kk,-} \approx -\sqrt{f^2_{0\kk,-}+f_{2\kk_g,-}\delta_{0}^2}  \approx -\sqrt{ v_1^2q_1^2+v_2^2q_2^2+ B^2_3\delta_{0}^2 }. 
\end{eqnarray}
This expression shows the linear dispersion (Dirac Hamiltonian) around the nodes if $\delta_{0}^2=0$, and $\delta_{0}^2$ plays the role of mass term for Dirac Hamiltonian. For its contribution to the condensation energy $E_c$, we need to evaluate the summation of $ E^-_{\kk,-}$ over the momentum $\qq$ in the region $\Omega_{\kk_g}$,  
\begin{eqnarray} 
&& \sum_{\kk \in \Omega_{\kk_g}} E^-_{\kk,-} = - \frac{S_M}{(2\pi)^2} \int_{\Omega_{\kk_g}} d^2q \sqrt{ v_1^2q_1^2+v_2^2q_2^2+ B_3 \delta_{0}^2} =  - \frac{S_M}{(2\pi)^2} \int_0^{\varepsilon} q dq \int_0^{2\pi} d\theta_q  \sqrt{ ( v_1^2 \cos^2\theta_q+v_2^2 \sin^2\theta_q) q^2 + B^2_3 \delta_{0}^2 } \nonumber \\
&& =  - \frac{S_M}{(2\pi)^2} \int_0^{2\pi} d\theta_q \frac{1}{3v_\theta^2}  \left( ( v_\theta^2 \varepsilon^2 + B^2_3 \delta_{0}^2 )^{3/2} - B_3^{3} \delta_0^3 \right)   
\end{eqnarray}
where $(q_1,q_2)=q(\cos\theta_q, \sin\theta_q)$, $S_M$ is the area of Moir\'e unit cell and $v_\theta^2=  v_1^2 \cos^2\theta_q + v_2^2 \sin^2\theta_q$. We note that $v_\theta^2>min(v_1^2, v_2^2)\doteq v_{min}^2$, so in our numerical calculations, we first choose the momentum cut-off $\varepsilon$ so that the linear dispersion (\ref{eq:eigen_energy_expansion1}) is valid for the momentum within the region $\kk \in \Omega_{\kk_g}$ and then choose the value of $\delta_0$ to be small enough to satisfy 
\beq \label{eq:nodalregion_condition} |B_3 \delta_0|\ll |v_{min} \varepsilon|\sim |\bar{v} \varepsilon|, \eneq
where $\bar{v}=(v_1+v_2)/2$. With this choice of the parameters $\varepsilon$ and $\delta_0$, we can perform the expansion
\begin{eqnarray}  \label{eq:EigenEng_negative_delta0}
&& \sum_{\kk \in \Omega_{\kk_g}} E^-_{\kk,-} =  - \frac{S_M}{(2\pi)^2} \int_0^{2\pi} d\theta_q  \frac{1}{3 v_\theta^2}  \left( ( v_\theta^2  \varepsilon^2 + B^2_3 \delta_{0}^2 )^{3/2} - B_3^{3} \delta_0^3 \right)   \approx - \frac{S_M}{3(2\pi)^2}  \int_0^{2\pi} d\theta_q  \left( v_\theta \varepsilon^3 \left(1 + \frac{3 B^2_3\delta_0^2}{2v_\theta^2\varepsilon^2}\right) - \frac{B_3^{3}}{v^2_\theta} \delta_0^3 \right) \nonumber \\
&& \approx - \frac{S_M}{(2\pi)^2}  \int_0^{2\pi} d\theta_q   \left(\frac{1}{3} v_\theta \varepsilon^3 + \frac{1}{2 v_\theta} B^2_3 \delta_0^2 \varepsilon \right) = - D_1 \varepsilon^3 - D_2 \varepsilon \delta_0^2
\end{eqnarray}
up to the order of $\delta_0^2$, 
where 
\begin{eqnarray}
    && D_1 = \frac{S_M}{(2\pi)^2}  \int_0^{2\pi} d\theta_q   \frac{1}{3} \sqrt{v_1^2 \cos^2\theta_q+v_2^2 \sin^2\theta_q}, \nonumber \\
    && D_2 = \frac{S_M}{(2\pi)^2}  \int_0^{2\pi} d\theta_q   \frac{B^2_3 }{2 \sqrt{v_1^2 \cos^2\theta_q+v_2^2 \sin^2\theta_q} } . 
\end{eqnarray}

From Eq. (\ref{eq:EigenEng_negative_delta0}), one can see that although non-zero $\delta_0$ can induce a gap that is proportional to $\delta_0$ (Eq. \ref{eq:EigenEng_gap}), the energy saving through the gap opening of the nodes is proportional to $\delta^2_{0}$ after taking into account the integral over the momentum in the whole $\Omega_{\kk_g}$ region. Thus, the $\delta_0$-dependence for the $\sum_{\kk \in \Omega_{\kk_g}}E^-_{\kk,n}$ and $\sum_{\kk \not\in \Omega_{\kk_g}} E^\pm_{\kk,n} $ terms is of the same order ($\delta_0^2$ order). As we do not find the linear-$\delta_0$ dependence for both $\sum_{\kk \in \Omega_{\kk_g}}E^-_{\kk,n}$ and $\sum_{\kk \not\in \Omega_{\kk_g}} E^\pm_{\kk,n} $, we conclude $\gamma=0$ in Eq. (\ref{eq:Ec_expansion}).  

We can make a further approximation and use the averaged velocity $\bar{v}=(v_1+v_2)/2$ to replace $v_1$ and $v_2$ in $D_{1,2}$, which leads to 
\begin{eqnarray} \label{eq:D1D2parameter}
    && D_1 \approx \frac{S_M}{(2\pi)^2} \frac{2\pi |\bar{v}|}{3} , \quad \quad  D_2 = \frac{S_M}{(2\pi)^2} \frac{\pi B^2_3 }{ |\bar{v}| } . 
\end{eqnarray}

\begin{figure}	\includegraphics[width=0.9\textwidth,angle=0]{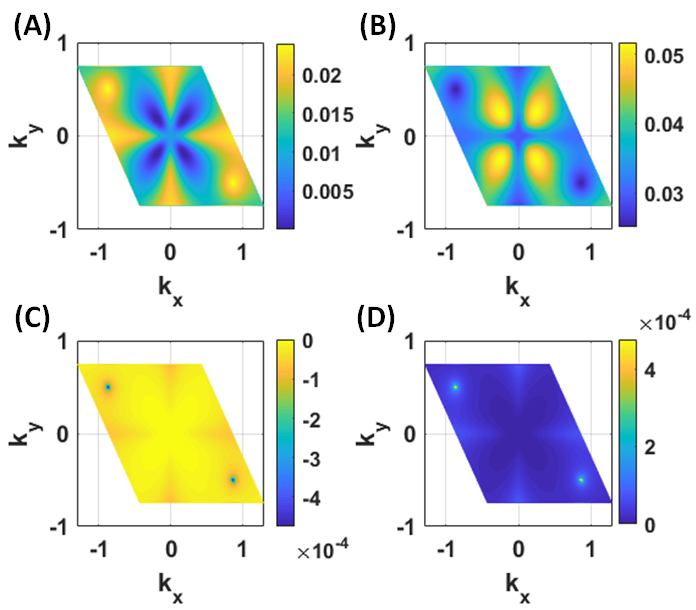}
	\centering
	\caption{ (A), (B), (C) and (D) show $f_{0\kk,-}^2 $, $f_{0\kk,+}^2$, $f_{2\kk,-} \delta_0^2$ and $f_{2\kk,+} \delta_0^2$ as a function of $\kk$ in the Moir\'e Brillouin zone, respectively. The unit of $f_{0\kk,n}^2$ and $f_{2\kk,n}\delta_0^2$ is meV$^2$.   }
	\label{fig_fparameter_comparison1}
\end{figure}

By substituting Eqs. (\ref{eq:EigenEng_negative_delta0}), (\ref{eq:Eng_expansion}) and (\ref{eq:Energy_MF1}) into Eq. (\ref{eq:condensation_energy_intra1}), we can obtain the expansion of $E_c$ for a small $\delta_0$ and $\phi=0$ in the form of Eq. (\ref{eq:Ec_expansion}) with 
\begin{eqnarray}
&& E_{c0} =  4 \left( - D_1 \varepsilon^3 - \sum_{n,\kk} \xi^-_{\kk,n} - \sum_{(n,\kk)\not\in (-,\Omega_{\kk_g})} f_{0\kk,n}  + \frac{ N_M}{V_0} \Delta^2_0  \right) \label{eq:condensation_expansion_1}  \\
&& \gamma =  0 \\
&& \beta =   4 \left( - D_2 \varepsilon - \sum_{(n,\kk)\not\in (-,\Omega_{\kk_g})} \frac{f_{2\kk,n}}{2f_{0\kk,n}} \right) . \label{eq:condensation_expansion_2} 
\end{eqnarray}
This is the main result of this section. One can see that the coefficient $\gamma=0$ of linear $\delta_0$ term is zero, so the Euler pairing with $\alpha=\pi/4$ is locally stable (unstable) if $\beta>0$ ($\beta<0$). We next analyze the sign of $\beta$ below in two steps. First, for $\delta_0=0$, we determine the value of $\Delta_0$ by minimizing $E_{c0}$. Then, we will analyze the sign of $\beta$, which determine if the Euler paring is a locally stable pairing or not. In Eqs. (\ref{eq:condensation_expansion_1}) and (\ref{eq:condensation_expansion_2}), we have the parameter $\varepsilon$, which is a tuning parameter to determine the node point regions ($\Omega_{\kk_g}$) and gapped regions in the momentum space. Its value is limited by the value of $\delta_0$ via the condition Eq. (\ref{eq:nodalregion_condition}). We can numerically test $E_{c0}$ and $\beta$ in Eqs. (\ref{eq:condensation_expansion_1}) and (\ref{eq:condensation_expansion_2}) for appropriate choice of the values of $\varepsilon$ and $\delta_0$ to satisfy the condition (\ref{eq:nodalregion_condition}). 
We first choose the momentum cut-off $\varepsilon \approx 0.01 k_\theta$, for which the area ratio between $\Omega_{\kk_g}$ and the MBZ is generally of order $10^{-4}$. Thus, $\Omega_{\kk_g}$ a very small region in the MBZ and the approximation of Dirac type of energy dispersion in Eq. (\ref{eq:eigen_energy_expansion1}) is well justified. With the above value of $\varepsilon$, we can safely choose $\delta_0=0.01\times \frac{\pi}{2}\approx 0.0157$, which is small enough to satisfy the condition (\ref{eq:nodalregion_condition}). With the numerical estimates of $\Delta_{\kk_g}\approx 0.14$ meV, $B_3=\left|\frac{\Delta_{0\kk_g}\mu}{\sqrt{2}d_{x,\kk_{g1}}}\right| \approx 0.036$meV, $\bar{v} \approx \frac{ 0.4 meV}{k_\theta}$, we find the condition (\ref{eq:nodalregion_condition}) becomes
\beq \varepsilon \gg \left|\frac{B_3 \delta_0}{\bar{v}}\right| \approx 0.0014 k_\theta, \eneq
which is well satisfied with our choice of $\varepsilon \approx 0.01 k_\theta$. 
This choice of $\delta_0$ is small enough for us to perform the perturbation expansion in Eq. (\ref{eq:Eng_expansion}). In Fig. \ref{fig_fparameter_comparison1}, we plot $f_{0\kk,-}^2 $, $f_{0\kk,+}^2$, $f_{2\kk,-} \delta_0^2$ and $f_{2\kk,+} \delta_0^2$ as a function of $\kk$ in the Moir\'e Brillouin zone, respectively, from which one can see the $f_{2\kk,n} \delta_0^2$ term is two orders smaller than $f_{0\kk,n}^2$, except around the momentum $\kk_g$ where $f_{0\kk_g,n}^2$ is close to zero (the region $\kk\in \Omega_{\kk_g}$). 

\begin{figure}	\includegraphics[width=0.9\textwidth,angle=0]{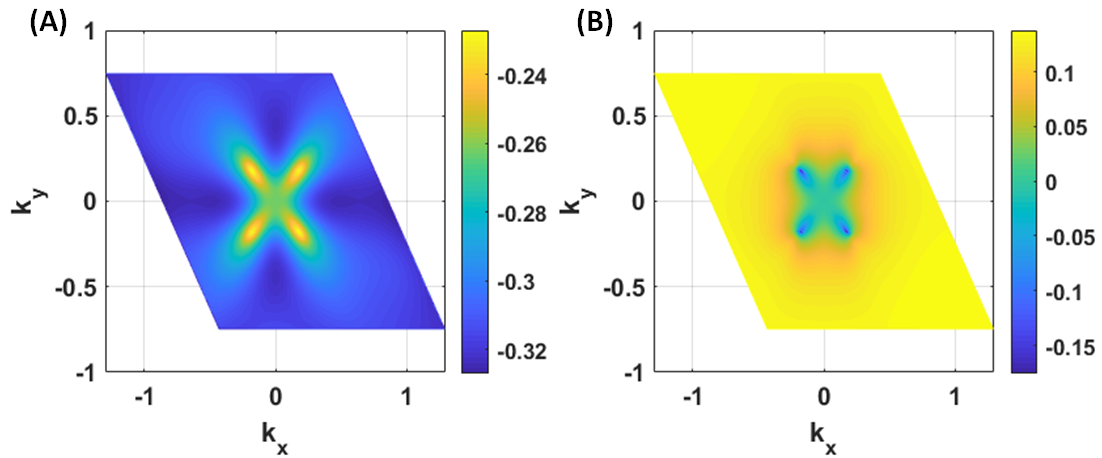}
	\centering
	\caption{ (A) shows $-\sum_{n} f_{0\kk,n}$ as a function of $\kk$ in the Moir\'e Brillouin zone.   (B) shows $-\sum_{n} \frac{f_{2\kk,n}}{2f_{0\kk,n}}$ as a function of $\kk$ in the Moir\'e Brillouin zone, which is directly proportional to $\beta_\kk$ defined in Eq. (\ref{eq:betak_1}). The energy unit here is meV.  }
	\label{fig_parameter_energy_1}
\end{figure}

With this choice of $\varepsilon$ value, we also estimate the terms $D_1\varepsilon^3 $ and $D_2 \varepsilon$,  
\begin{eqnarray}
\label{eq:D1D2_estimatevalue}
& D_1\varepsilon^3 \approx  \frac{2\pi \varepsilon^2 S_M}{(2\pi)^2} \frac{\bar{v}\varepsilon}{3} \approx  3\times 10^{-7} meV; \nonumber \\
& D_2\varepsilon \approx \frac{\pi \varepsilon^2 S_M}{(2\pi)^2} \frac{\pi B^2_3 }{ \bar{v} \varepsilon }  \approx 4\times 10^{-5} meV. 
\end{eqnarray}
Here the factor $\frac{\pi \varepsilon^2 S_M}{(2\pi)^2}$ represents the area ratio between $\Omega_{\kk_g}$ and the MBZ, and is generally of order $10^{-4}$, which is thus a very small region in the MBZ. As a comparison, Fig. \ref{fig_parameter_energy_1} (A) and (B) shows $-\sum_{n} f_{0\kk,n}$ and $-\sum_{n} \frac{f_{2\kk,n}}{2f_{0\kk,n}}$, both of which is of order $10^{-1}$ meV. Thus, we conclude that by choosing $\delta_0=0.01\times \frac{\pi}{2}$ and $\varepsilon \approx 0.01 k_\theta$, we can satisfy the condition (\ref{eq:nodalregion_condition}) and safely drop the term $D_1\varepsilon^3 $ and $D_2\varepsilon$ in calculating $E_{c0}$ (Eq. \ref{eq:condensation_expansion_1}) and $\beta$ (Eq. \ref{eq:condensation_expansion_2}).

Indeed, our numerical calculation gives $\beta \approx 0.44$meV, which is a positive number and much larger than the $D_2\varepsilon$ term in Eq.(\ref{eq:D1D2_estimatevalue}). We next analyze the sign of the dominant term in $\beta$ (Eq.\ref{eq:condensation_expansion_2}) after dropping the first term $D_1\varepsilon^3$ in $E_{c0}$. One can show
\beq \frac{\partial f_{0\kk,n}}{\partial \Delta_0}=\frac{f_{1\kk,n}}{2f_{0\kk,n}} \frac{\partial \Delta_{0\kk}}{\partial \Delta_0}
=\frac{f_{1\kk,n}}{2f_{0\kk,n}} \frac{k^2}{k^2+b^2}, 
\eneq
where 
\beq f_{1\kk,n}=\Delta_{0\kk} \left( 1 + n\frac{|d_{x,\kk}|\sin^2(2\theta_\kk)}{\sqrt{\mu^2+2\Delta_{0\kk}^2\sin^2(2\theta_\kk)}} \right),  \eneq
and thus 
\begin{eqnarray} && \frac{\partial E_{c0}}{\partial \Delta_0}=0 \rightarrow  - \sum_{(n,\kk)\not\in (-,\Omega_{\kk_g})} \frac{\partial f_{0\kk,n}}{\partial \Delta_0} + \frac{2 N_M}{V_0} \Delta_0 =0 \rightarrow  \nonumber\\
&& \frac{2 N_M}{V_0} \Delta_0 = \sum_{(n,\kk)\not\in (-,\Omega_{\kk_g})} \frac{f_{1\kk,n}}{2f_{0\kk,n}} \frac{k^2}{k^2+b^2}  \label{eq:gapequation_Delta0} .  
\end{eqnarray}
This is just the self-consistent equation that determines the value of $\Delta_0$, which minimizes $E_{c0}$. 

After dropping the first term $D_2\varepsilon$ in $\beta$, the second term of $\beta$ in Eq.(\ref{eq:condensation_expansion_2}) can be written as 
\begin{eqnarray}\label{eq:beta_full1}
&& \beta = - 2 \sum_{(n,\kk)\not\in (-,\Omega_{\kk_g})} \frac{f_{2\kk,n}}{f_{0\kk,n}} = - \sum_{\kk\not\in \Omega_{\kk_g}} \sum_{n=\pm} \frac{n}{f_{0\kk,n}}  \frac{2\Delta_{0\kk}^4 + \Delta_{0\kk}^2 d_{x,\kk}^2\cos(4\theta_\kk)}{|d_{x,\kk}|\sqrt{\mu^2+2\Delta_{0\kk}^2\sin^2(2\theta_\kk)}}, 
\end{eqnarray}
where we have used the expression (\ref{eq:fun2kn}) for $f_{2\kk,n}$. 

The numerical results for $-\sum_{n} \frac{f_{2\kk,n}}{2f_{0\kk,n}}$ as a function of the momentum $\kk$ is numerically shown in Fig. \ref{fig_parameter_energy_1}(B), from which one can see $-\sum_{n} \frac{f_{2\kk,n}}{2f_{0\kk,n}}$ is mostly positive value in the MBZ and its value becomes negative around the point nodes $\kk_g$. To understand the numerical results, we denote 
\beq \label{eq:betak_1} \beta = \sum_{\kk\not\in \Omega_{\kk_g}} \beta_{\kk}; \quad \quad  \beta_{\kk} = - 2 \sum_n  \frac{f_{2\kk,n}}{f_{0\kk,n}} =  - \left(\sum_n \frac{n}{f_{0\kk,n}} \right) \frac{2\Delta_{0\kk}^4 + \Delta_{0\kk}^2 d_{x,\kk}^2\cos(4\theta_\kk)}{|d_{x,\kk}|\sqrt{\mu^2+2\Delta_{0\kk}^2\sin^2(2\theta_\kk)}}. \eneq

$\beta_{\kk}$ has two parts: a summation over $n$-dependent part $- \left(\sum_n \frac{n}{f_{0\kk,n}} \right) $ and a $n$-independent part $\frac{2\Delta_{0\kk}^4 + \Delta_{0\kk}^2 d_{x,\kk}^2\cos(4\theta_\kk)}{|d_{x,\kk}|\sqrt{\mu^2+2\Delta_{0\kk}^2\sin^2(2\theta_\kk)}}$, where $n=\pm$. 
From the expression of $f_{0\kk,n}$ (Eq. (\ref{eq:f0_eigenenergy})), $f_{0\kk,+}>f_{0\kk,-}>0$. Thus, $-\left(\sum_n\frac{n}{f_{0\kk,n}}\right)=-\frac{1}{f_{0\kk,+}}+\frac{1}{f_{0\kk,-}}>0$. 
As $|d_{x,\kk}|\sqrt{\mu^2+2\Delta_{0\kk}^2\sin^2(2\theta_\kk)}$ is positive, together with $-\left(\sum_n\frac{n}{f_{0\kk,n}}\right)=-\frac{1}{f_{0\kk,+}}+\frac{1}{f_{0\kk,-}}>0$, the sign of $\beta_{\kk}$ is then completely determined by the sign of   
$2\Delta_{0\kk}^4 + \Delta_{0\kk}^2 d_{x,\kk}^2\cos(4\theta_\kk)$. 
For the momentum $\kk$ with the angle around $\theta_\kk \approx 0, \pm \frac{\pi}{2}, \pi$,  $\cos4\theta_\kk\approx +1 >0$, 
all the terms in $2\Delta_{0\kk}^4 + \Delta_{0\kk}^2 d_{x,\kk}^2\cos(4\theta_\kk) \approx 2\Delta_{0\kk}^4 + \Delta_{0\kk}^2 d_{x,\kk}^2$ are positive, so we find $\beta_\kk$ to be positive. 
For the momentum $\kk$ with the angle around $\theta_\kk \approx \pm\frac{\pi}{4}, \pm\frac{3\pi}{4}$, we find $\cos(4\theta_\kk)\approx -1$ and  
\beq 2\Delta_{0\kk}^4 + \Delta_{0\kk}^2 d_{x,\kk}^2\cos(4\theta_\kk) \approx 2\Delta_{0\kk}^4 - \Delta_{0\kk}^2 d_{x,\kk}^2 = \Delta_{0\kk}^2 \left( 2\Delta_0^2 \frac{k^2}{k^2+b^2} - d_{x,\kk}^2 \right), \eneq 
the sign of which depends on the amplitudes of two terms $2\Delta_{0\kk}^4$ and $\Delta_{0\kk}^2 d_{x,\kk}^2$. We find the following features: (1)  
Around $\KK_M$, since $d_{x,\kk=\KK_M}\approx 0$, we have $\Delta_{0 \kk=\KK_M}^2\gg d_{x,\kk=\KK_M}^2$, so we expect $\beta_\kk>0$. From numerical calculations, we generally find $d_{x,\kk}$ to be small for a large momentum $\kk$ around MBZ boundary, and thus $\beta_\kk>0$ is found to be positive for a large momentum (Fig. \ref{fig_parameter_energy_1}(B)). (2) On the other hand, for a small momentum $|\kk|\ll b$, $\Delta_{0\kk}=\Delta_0 \frac{k^2}{k^2+b^2} \rightarrow 0$, so $\beta_\kk < 0$ for the small momentum $\kk$ around the momentum angle $\theta_\kk \approx \pm\frac{\pi}{4}, \pm\frac{3\pi}{4}$. (3) Around the point nodes $\kk\approx \kk_g$, we have 
\beq 2\Delta_{0\kk_g}^4 + \Delta_{0\kk_g}^2 d_{x,\kk_g}^2\cos(4\theta_{\kk_g}) \approx \Delta_{0\kk_g}^2 ( 2\Delta_{0\kk_g}^2 - d_{x,\kk_g}^2) = -  \Delta_{0\kk_g}^2 \tilde{d}_{0,\kk_g}^2 <0 \eneq 
where we have used the condition (\ref{eq:gapless_condition}). Thus, $\beta_\kk < 0$ around the point nodes. All these features are consistent with our numerical results in Fig. \ref{fig_parameter_energy_1}(B). Since $\beta_\kk$ is positive for most momenta from Fig. \ref{fig_parameter_energy_1}(B), the summation of $\beta_\kk$ over the momentum space gives a positive $\beta$, which is around $\beta\sim 0.44$ meV from our numerical calculations. 

Finally, we discuss the expansion of $E_c$ in Eq.(\ref{eq:Ec_expansion}) for other values of $\phi$. For a fixed value of $\phi$, according to $\Phi_\kk=\phi+4\theta_\kk$ and Eq. (\ref{eq:nodeAngle1}), the momentum angle of nodes should satisfy 
\beq \theta_{\kk_g}=\pm\frac{\pi}{4}-\frac{\phi}{4}, \quad \pm \frac{3\pi}{4}-\frac{\phi}{4}. \eneq 
Thus, the node momentum $\kk_g$ can be obtained by rotating the momentum vectors $\frac{k_0}{\sqrt{2}}(\pm 1, \pm 1)$ by an angle $\phi/4$. We can generally make a coordinate transformation in the momentum space by rotating the momentum by an angle $\phi/4$, defined in Eq.(\ref{eq:coordinateTransformation1}). In the new coordinate, we will have $\Phi_{\kk'}=4\theta_{\kk'}$ and $\theta_{\kk'_g}=\pm\frac{\pi}{4}, \quad \pm \frac{3\pi}{4}$, so that the nodes are located at 
$\kk'_{g1}=\frac{k_0}{\sqrt{2}}(1,1)$, $\kk'_{g2}=\frac{k_0}{\sqrt{2}}(1,-1)$, $\kk'_{g3}=\frac{k_0}{\sqrt{2}}(-1,1)$ and $\kk'_{g4}=\frac{k_0}{\sqrt{2}}(-1,-1)$ in the new coordinate. All the expansion can be performed in the new coordinate with a similar form. As our numerical results have demonstrated the weak dependence on $\phi$, we would expect to get a similar value of $\beta$. 

\section{Importance of inter-eigen-band pairing}
In this section, we will transform the BdG Hamiltonian from Chern-band basis to the eigen-state basis and will show the importance of the inter-eigen-band pairing in stablizing the gapless nematic SC phase. To do that, we first apply a unitary transformation
\beq 
U_0=\frac{1}{\sqrt{2}}
\begin{pmatrix}
1 & -1 & 0 & 0 \\
1 & 1 & 0 & 0 \\
0 & 0 & 1 & -1 \\
0 & 0 & 1 & 1
\end{pmatrix}
\eneq
to the BdG Hamiltonian (\ref{eq:HBdGpup}) and transforms the single-particle Hamiltonian part of $\mathcal{H}_{BdG}^{+,+}$ into a diagonal form. After the transformation, we have
\beq 
\tilde{\mathcal{H}}_{BdG}^{+,+}=U_0^\dagger\mathcal{H}_{BdG}^{+,\uparrow}U_0=
\begin{pmatrix}
\epsilon_{\kk,+}/2 & 0 & (\Delta_{+,\kk}+\Delta_{-,\kk})/2 &  (\Delta_{-,\kk}-\Delta_{+,\kk})/2 \\
0 & \epsilon_{\kk,-}/2 & (\Delta_{-,\kk}-\Delta_{+,\kk})/2 & (\Delta_{+,\kk}+\Delta_{-,\kk})/2 \\
(\Delta^\star_{+,\kk}+\Delta^\star_{-,\kk})/2 & (\Delta^\star_{-,\kk}-\Delta^\star_{+,\kk})/2 & - \epsilon_{\kk,+}/2 & 0 \\
 (\Delta^\star_{-,\kk}-\Delta^\star_{+,\kk})/2 & (\Delta^\star_{+,\kk}+\Delta^\star_{-,\kk})/2 & 0 & - \epsilon_{\kk,-}/2
\end{pmatrix},
\eneq
where $\epsilon_{\kk,\pm}=d_{0,\kk}-\mu\pm d_{x,\kk}$ and the gap function $\Delta_{\kk,e_Y}$ is taken the d-wave ansatz in Eq. (\ref{eq:dwave_ansatz0}). 
We notice that in this eigen-state basis, the intra-Chern-band pairing not only occurs within each eigen-energy band (intra-eigen-band pairing), 
but also can exist between two eigen-energy bands (inter-eigen-band pairing). We will next argue that the inter-eigen-band pairing is essential in stablizing the Euler pairing. 

In the previous section, we have shown that the $\beta$ in the expansion (\ref{eq:Ec_expansion}) of the condensation energy $E_c$ has a positive sign, so that the Euler pairing is locally stable. Now we manually turn off the inter-eigen-band pairing and consider the Hamiltonian without inter-eigen-band pairing terms
\beq \label{eq:HBdG_comparison}
\tilde{H}'_{BdG}=
\begin{pmatrix}
\epsilon_{\kk,+}/2 & 0 & (\Delta_{\kk,+}+\Delta_{\kk,-})/2 & 0 \\
0 & \epsilon_{\kk,-}/2 & 0 & (\Delta_{\kk,+}+\Delta_{\kk,-})/2 \\
(\Delta^\star_{\kk,+}+\Delta^\star_{\kk,-})/2 & 0 & - \epsilon_{\kk,+}/2 & 0 \\
0 & (\Delta^\star_{\kk,+}+\Delta^\star_{\kk,-})/2 & 0 & - \epsilon_{\kk,-}/2
\end{pmatrix}. 
\eneq
and will calculate the condensation energy $E_c$ from this Hamiltonian. For the numerical calculations in this section, we increase the interaction parameter $V_0=0.8 meV$ in order to stablize the SC phase, while all the other parameters are kept the same. 

\begin{figure}
	\includegraphics[width=0.9\textwidth,angle=0]{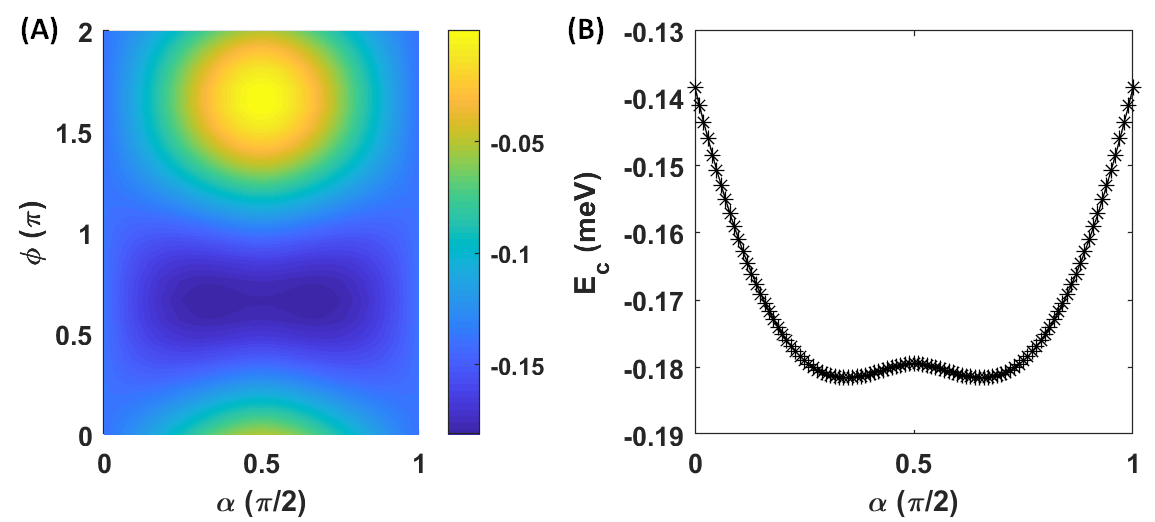}
	\centering
	\caption{ (A) The condensation energy for the BdG Hamiltonian (\ref{eq:HBdG_comparison}) as a function of $\alpha$ and $\phi$ where $\Delta_0$ is chosen to be 0.29 meV from the self-consistent gap equation. The red dashed line corresponds to the $\phi$ value for the plot in (B). (B) $E_c$ as a function of $\alpha$ for $\phi\approx 0.6667\pi$.  }
	\label{fig_Ec_alp_phi_intra_eigen}
\end{figure}

The negative BdG eigen-energies of the Hamiltonian (\ref{eq:HBdG_comparison}) are given by 
\begin{eqnarray} \label{eq:BdGEnergy2}
& E^{'-}_{\kk,n}= - \frac{1}{2}\sqrt{( |d_{x,\kk}|+n |\tilde{d}_{0,\kk}|)^2 +|\Delta_{\kk,-}|^2+|\Delta_{\kk,+}|^2+2|\Delta_{\kk,+}\Delta_{\kk,-}|
\cos(\Phi_{\kk}) },
\end{eqnarray}
where $n=\pm$. From Eq. (\ref{eq:phik_gapfunction}),  $\Phi_\kk$ depends on the relative phase $\phi$ and momentum angle $\theta_\kk$. Using the BdG eigen-energy in Eq.(\ref{eq:BdGEnergy2}) and the gap function ansatz Eq.(\ref{eq:gap_function_expansion}), the condensation energy $E_c$ can be numerically evaluated from Eq. (\ref{eq:condensation_energy_intra1}). Fig. \ref{fig_Ec_alp_phi_intra_eigen}(A) shows $E_c$ as a function of $\alpha$ and $\phi$. The $E_c$ from BdG eigen-energy (\ref{eq:BdGEnergy2}) has a strong dependence on the relative phase $\phi$ between the gap function $\Delta_+$ and $\Delta_-$, which is quite different from the weak $\phi$ dependence of $E_c$ in Fig. \ref{fig_condeng_comp_3}(B) with the full pairing form. The minimal $E_c$ is found around $\phi=\phi_0\approx 0.6667 \pi$, which originates from the in-plane anisotropic energy spectrum of the flat bands (or equivalently the in-plane anisotropy of the function $d_{x,\kk}$), as discussed above in Sec. B. We show $E_c$ as a function of $\alpha$ for $\phi=\phi_0\approx 0.6667 \pi$ in Fig. \ref{fig_Ec_alp_phi_intra_eigen}(B). We notice that the minimal $E_c$ is located at $\alpha\approx 0.34\times \frac{\pi}{2}$, away from the Euler pairing with $\alpha=\pi/4$.  

To understand why the Euler pairing becomes unstable in this case, we consider an expansion of the $E_c$ around $\alpha=\pi/4+\delta_0$ with $\delta_0\ll 1$. The negative eigen-energies are 
\begin{eqnarray} \label{eq:BdG_eigen_energy2}
& E^{'-}_{\kk,n}= - \frac{1}{2}\sqrt{(|d_{x,\kk}| + n |\mu|)^2 + \Delta_{0\kk}^2 (1+\cos(\Phi_\kk)) - 2\Delta_{0\kk}^2 \cos(\Phi_\kk) \delta_0^2 },
\end{eqnarray}
where we have used $\tilde{d}_{0,\kk}=-\mu$. We also notice that for the Euler paring ($\delta_0=0$), $E^{'-}_{\kk,-}$ has the nodes at $\Phi_\kk =\pm\pi, \pm 3\pi$ so that $\cos(\Phi_\kk)=-1$, and $|d_{x,\kk}|=|\mu|$. As $\Phi_\kk=\phi+4\theta_\kk$, the momentum angle of nodes should satisfy $\theta_\kk=\frac{\pm\pi-\phi}{4}, \frac{\pm 3\pi-\phi}{4}$. Thus, the node momentum $\kk_g$ can be obtained by rotating the momentum vectors $\frac{k_0}{\sqrt{2}}(\pm 1, \pm 1)$ by an angle $\phi/4$, in which $k_0=\sqrt{(C_0-|\mu|)/C_2}$ with the approximation $d_{x,\kk}\approx C_0-C_2 k^2$ around $\Gamma_M$. 

The expansion of the eigen-energy around the nodes is similar to that in Sec. \ref{Sec:CondenstationEnergy}. We can separate the Moir\'e BZ into two regions, a small region around nodes at $\kk_g$ with the cut-off $\varepsilon$ (labelled by $\Omega_{\kk_g}$) and the region away from the nodes. Around the nodes $\kk\in \Omega_{\kk_g}$, one can show the eigen-energy can be expanded as
\begin{eqnarray} 
& E^{'-}_{\kk,n}= - \sqrt{ v_1^2q_1^2 + v_2^2 q_2^2 + B_3^2 \delta_0^2 },
\end{eqnarray}
where $\kk=\kk_g+\qq$ and 
\beq q_1=(q_x+q_y)/\sqrt{2}, \; q_2=(q_x-q_y)/\sqrt{2}, \; v_1=C_2k_0, \;  v_2=\sqrt{2}\frac{\Delta_{0\kk_g}}{k_0}, \; B_3=\frac{1}{\sqrt{2}} \Delta_{0\kk_g}. 
\eneq 
The above expression is the same as Eq. (\ref{eq:eigen_energy_expansion1}), and thus we can perform the integral over $\qq$ and expand the expression for a small $\delta_0$,  
\begin{eqnarray} 
&& \sum_{\kk \in \Omega_{\kk_g}} E^{'-}_{\kk,-} \approx - D_1 \varepsilon^3 - D_2 \varepsilon \delta_0^2, 
\end{eqnarray}
where $D_1$ and $D_2$ are given by Eqs. (\ref{eq:D1D2parameter}). The expansion of the condensation energy is still given by Eq.(\ref{eq:Ec_expansion}) with $E_{c0}$ determined by Eq. (\ref{eq:condensation_expansion_1}), $\gamma=0$ and $\beta$ given by Eq. (\ref{eq:condensation_expansion_2}). 

We again use the average velocity $\bar{v}$ obtained from our numerical results for the estimate $v_1$ and $v_2$, which is around $\bar{v}\sim \frac{0.6}{k_\theta} meV$. We further estimate $B_3\sim 0.16 meV$, so that $\varepsilon \gg 0.0043 k_\theta$ from the condition (\ref{eq:nodalregion_condition}). Thus, we can choose $\varepsilon \sim 0.05 k_\theta$ to satisfy this condition, and find 
\begin{eqnarray}
& D_1\varepsilon^3 \approx  \frac{2\pi \varepsilon^2 S_M}{(2\pi)^2} \frac{\bar{v}\varepsilon}{3} \approx  6\times 10^{-5} meV; \nonumber \\
& D_2\varepsilon \approx \frac{\pi \varepsilon^2 S_M}{(2\pi)^2} \frac{\pi B^2_3 }{ \bar{v} \varepsilon }  \approx 0.0026 meV, 
\end{eqnarray}
both of which are also negligible.

\begin{figure}
	\includegraphics[width=0.9\textwidth,angle=0]{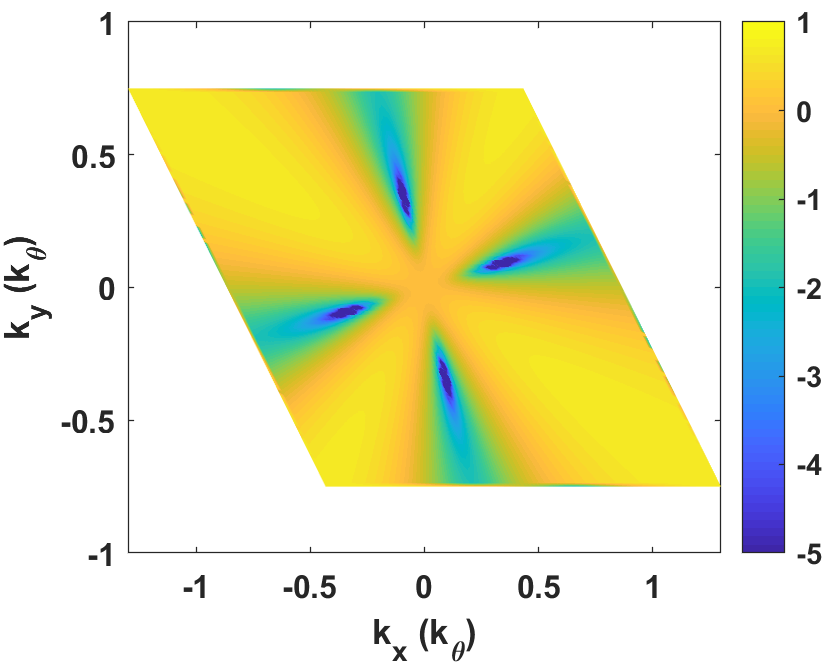}
	\centering
	\caption{  $\beta_\kk$ as a function of the momentum $\kk$. The integral of $\beta_{\kk}$ in the MBZ is around $\beta \sim -0.12 meV$, which has a negative sign. }
	\label{fig_beta_k_intra_eigen}
\end{figure}

For $\kk\not\in \Omega_{\kk_g}$, we can perform the perturbation expansion. We define 
\beq f'_{0\kk,n}=\frac{1}{2}\sqrt{(|d_{x,\kk}| +n |\mu|)^2 + \Delta_{0\kk}^2 (1+\cos(\Phi_\kk))}
\eneq
and
\beq f'_{2\kk,n}=-\frac{1}{2}\Delta^2_{0\kk} \cos(\Phi_\kk),
\eneq
so that
\beq E^{'-}_{\kk,n}=-\sqrt{(f'_{0\kk,n})^2 +  f'_{2\kk,n}\delta_0^2}
\approx - f'_{0\kk,n} \left( 1 + \frac{f'_{2\kk,n}}{2(f'_{0\kk,n})^2} \delta_0^2 \right). 
\eneq
With all the above calculation, one can show that the expansion of the condensation energy $E_c$ in Eq. (\ref{eq:Ec_expansion}) has the coefficient, 
\begin{eqnarray}
&& E_{c0} =  4 \left( - D_1 \varepsilon^3 - \sum_{n,\kk} \xi^-_{\kk,n} - \sum_{(n,\kk)\not\in (-,\Omega_{\kk_g})} f'_{0\kk,n}  + \frac{ N_M}{V_0} \Delta^2_0  \right)  \\
&& \gamma =  0 \\
&& \beta =   4 \left( - D_2 \varepsilon - \sum_{(n,\kk)\not\in (-,\Omega_{\kk_g})} \frac{f'_{2\kk,n}}{2f'_{0\kk,n}} \right) =   4 \left( - D_2 \varepsilon + \sum_{(n,\kk)\not\in (-,\Omega_{\kk_g})} \frac{\Delta_{0\kk}^2}{4f'_{0\kk,n}} \cos \Phi_{\kk} \right) . 
\end{eqnarray}
The most important is to determine the sign of $\beta$. As $D_2\varepsilon$ term is negligible, we define 
\beq \beta_\kk = \sum_n \frac{\Delta_{0\kk}^2}{f'_{0\kk,n}} \cos \Phi_{\kk},  \eneq
and Fig. \ref{fig_beta_k_intra_eigen} shows the distribution of $\beta_\kk$ in the Moir\'e BZ. 

When $\theta_\kk$ is close to $\frac{\pm\pi-\phi_0}{4}, \frac{\pm3\pi-\phi_0}{4}$, $\cos(\Phi_\kk)\approx -1$, so the contribution to $\beta_\kk$ around this momentum angle is 
\beq \sum_{n} \frac{\Delta_{0\kk}^2}{f'_{0\kk,n}}\cos(\Phi_\kk) \approx - \sum_n \frac{2\Delta_{0\kk}^2}{\left||d_{x,\kk}| +n |\mu|\right|} < 0.   
\eneq 
Indeed, one can see a strong negative contribution to $\beta_\kk$ around the node region in Fig. \ref{fig_beta_k_intra_eigen}. On the other hand, when $\theta_\kk$ is close to $\pm\frac{\pi}{2}-\frac{\phi_0}{4}, -\frac{\phi_0}{4}, \pi-\frac{\phi_0}{4}$, $\cos(\Phi_\kk)\approx +1$ and the contribution to $\beta_\kk$ around this momentum angle is 
\beq \sum_{n} \frac{\Delta_{0\kk}^2}{f'_{0\kk,n}}\cos(\Phi_\kk) \approx  \sum_{n} \frac{2\Delta_{0\kk}^2}{\sqrt{(|d_{x,\kk}| +n |\mu|)^2 + 2 \Delta_{0\kk}^2}} > 0.
\eneq 
The magnitude of the contribution around $\theta_\kk \sim \frac{\pm\pi-\phi_0}{4}, \frac{\pm3\pi-\phi_0}{4}$ is found to be larger than that from $\theta_\kk \sim \pm\frac{\pi}{2}-\frac{\phi_0}{4}, -\frac{\phi_0}{4}, \pi-\frac{\phi_0}{4}$, and numerical results show $ \sum_{\kk\not\in \Omega_{\kk_g}} \beta_{\kk} \approx -0.12 meV < 0$. Thus, without inter-eigen-band pairing, the gapless nematic phase becomes unstable. This implies the importance of inter-eigen-band pairing in stablizing
the gapless nematic SC phase. We note that the inter-eigen-band pairing is a consequence of strong coupling nature of the multiple flat bands.

\end{document}